\documentclass[prb, twocolumn]{revtex4}
\usepackage{graphicx}
\usepackage{amsmath}

\usepackage{amsmath}
\usepackage{amsfonts}

\usepackage{epsfig} 

\usepackage{epstopdf}
\newcommand{\Ket}[1]{\left|#1  \right>}

\newcommand{\Braket}[1]{\left<#1  \right>}

\def\begeq{\begin{equation}}
\def\endeq{\end{equation}}

\def\begeqar{\begin{eqnarray}}
\def\endeqar{\end{eqnarray}}

\usepackage{color}

\begin{document}

\title{Logarithmic Correlations in Quantum Hall Plateau Transitions}

\author{Romain Vasseur}

\affiliation{Department of Physics, University of California, Berkeley, Berkeley CA 94720, USA}
\affiliation{Materials Science Division, Lawrence Berkeley National Laboratory, Berkeley CA 94720, USA}
\date{\today}

\begin{abstract}

The critical behavior of quantum Hall transitions in two-dimensional disordered electronic systems can be described by a class of complicated, non-unitary conformal field theories with logarithmic correlations. The nature and the physical origin of these logarithmic correlation functions remain however mysterious. Using the replica trick and the underlying symmetries of these quantum critical points, we show here how to construct non-perturbatively disorder-averaged observables in terms of Green's functions that scale logarithmically at criticality. In the case of the spin quantum Hall transition, which may occur in disordered superconductors with spin-rotation symmetry and broken time reversal invariance, we argue that our results are compatible with an alternative approach based on supersymmetry. The generalization to the Integer quantum Hall plateau transition is also discussed. 

\end{abstract}

\maketitle

\section{Introduction}

Random impurities in systems of non-interacting electrons can induce a transition between metallic (delocalized) and insulating (localized) phases~\cite{PhysRev.109.1492}. More than fifty years after its discovery, the field of Anderson localization remains very active~\cite{abrahams201050}. Even though the standard scaling theory of localization~\cite{PhysRevLett.42.673} predicts  the absence of extended states in disordered non-interacting electronic systems in two dimensions, a well-known exception to this rule is provided by the transition between plateaux in the Integer Quantum Hall Effect (IQHE), a quantum critical phenomena that was predicted theoretically and observed experimentally a few decades ago~\cite{PhysRevLett.45.494, PhysRevLett.48.1559,PhysRevLett.61.1294}.

The physics of non-interacting electrons moving on a two-dimensional plane with a perpendicular magnetic field and a random potential can be described by a two-dimensional sigma model on the manifold $M_n=U(2n)/U(n)\times U(n)$ in the limit $n \to 0$~\cite{Pruisken:1984aa}. The sigma model can be argued to be asymptotically free, and it flows to strong coupling, suggesting complete localization in dimension $d=2$ (this is related to the absence of Goldstone phase in $d=2$). The reason why the IQHE transition is allowed is well understood in terms of an additional topological $\theta$ term -- that can be written because $\pi_2(M_n)=\mathbb{Z}$ -- in the sigma model that induces a transition at $\theta=\pi$. It is by now well-admitted that the quantum critical point corresponding to this IQHE transition -- the strong coupling fixed point of the sigma model at $\theta=\pi$ -- should be effectively described by a Conformal Field Theory (CFT) in 1+1 dimensions (see {\it e.g.} Refs.~\onlinecite{1999hep.th5054Z,PhysRevLett.112.186803}). 

This effective description of a disordered electronic system in 2+1d in terms of a CFT in 1+1d comes at the price of losing unitarity, and as a result, the critical properties of the IQHE transition remain quite poorly understood: the critical exponents are known only either from experiments or numerical simulations, and a low energy field theory description of the critical point is still missing. Because of the underlying replica or supersymmetric description of the disordered critical point, it is however easy to argue that the central charge of this CFT should be $c=0$~\cite{Gurarie1993535,2004hep.th9105G} -- because the partition function of the problem is essentially trivial $Z=1$ -- 
and as such should be a Logarithmic Conformal Field Theory (LCFT)~\cite{Gurarie1993535,2004hep.th9105G,0305-4470-35-27-101, 1751-8121-46-49-494001, 1751-8121-46-49-494003} (the only unitary non-logarithmic CFT at $c=0$ being trivial). LCFTs are characterized by the non-diagonalizability of the scale transformation generator, corresponding to {\it indecomposable} (reducible but not fully reducible), non-unitary representations of the Virasoro algebra. This property leads to the appearance of logarithms in correlation functions and to ``indecomposable'' operator product expansions. These quantum field theories are daunting for many reasons: they are necessarily non-unitary, they are typically non-rational ({\it i.e.} they involve infinitely many primary fields), and from an algebraic point of view they require going beyond the description in terms of irreducible representations familiar to physicists, and are instead built out of large, complicated indecomposable representations that are very hard to classify mathematically~\cite{WildGermoni}. Most tools familiar to physicists in this context also have to be reconsidered: for instance the Mermin-Wagner theorem does not apply, and conformal invariance and global group symmetry does not imply Kac Moody symmetry. For example, whereas the $O(3)$ sigma model~\cite{PhysRevLett.50.1153} on the coset $S^2=SU(2)/U(1)$ at topological angle $\theta=\pi$ flows to a Wess-Zumino-Witten (WZW) CFT with enlarged symmetry $SU(2)_1$ that is ``easily'' manageable~\cite{PhysRevB.36.5291}, the CFT describing the IQHE transition (also obtained from a seemingly similar sigma model) is most likely not a WZW model. Despite the recent progress in the understanding of such LCFTs both from lattice models~\cite{Read2007316,1742-5468-2006-11-P11017,ReviewLCFTlattice} and more abstract algebraic approaches (see {\it e.g.} Refs.~\onlinecite{1742-5468-2006-04-P04002,0305-4470-39-49-012,Mathieu2007120,Mathieu2008268,kytola123503,1751-8121-42-32-325403}), all examples of LCFTs that are slowly getting under control are still pretty far from the expected complexity of the LCFT describing the IQHE plateau transition. 

Even if a detailed understanding of this logarithmic CFT seems unfortunately out of reach for now, it should be possible to understand what {\it physical observables} make the theory logarithmic. One of the main features of a LCFT is indeed the existence of logarithmic correlations at criticality -- where one usually expects only power-laws; it is therefore very natural to ask what kind of disorder-averaged observables in the IQHE show logarithmic correlations at the plateau transition. A general mechanism was proposed by Cardy~\cite{1999cond.mat.11024C,1751-8121-46-49-494001} to explain the emergence of logarithmic correlations in disordered systems from the replica trick, but this argument has to be adapted to the more complicated symmetry structure of the IQHE critical point. This approach was remarkably successful to understand the physical origin of logarithmic correlations in ``simpler'' problems including the $O(n \to 0)$ model~\cite{1999cond.mat.11024C,1751-8121-46-49-494001}, or percolation~\cite{1742-5468-2012-07-L07001} for example. Even though the IQHE transition is admittedly much more involved, it is still natural to expect the (heuristic) arguments of Refs.~\onlinecite{1999cond.mat.11024C,1751-8121-46-49-494001} to shed some light on the physical origins of logarithmic correlations in Anderson transitions. 
 
Following Cardy, our general strategy in this paper will be to use the replica trick to express disorder-averaged observables (Green's functions) in terms of correlators of $n$ species of fermionic operators, with $n$ the number of replicas, in a theory with a global group symmetry $G(n)$ -- typically the unitary group $U(n)$. Using the representation theory of this global symmetry group, we classify the different lattice observables and deduce the general form of their correlation functions using symmetry arguments only. The critical properties of the physical, disordered system can then be understood, in principle, from the formal $n \to 0$ limit of the critical point (if it exists) of these replicated theories for generic $n$ integer. The resulting correlation functions usually have $1/n$ poles and are therefore ill-defined in the physical limit $n \to 0$. These apparent divergences in the limit $n \to 0$ can be cured but they lead to logarithms in the limit. The basic mechanism underlying the emergence of logarithms in the replica limit is quite easily understood: starting from two power-law functions $r^{-2 \Delta_{1}(n)} / n$ and $r^{-2 \Delta_{2}(n)}/n$ with coinciding critical exponents for $n=0$, both diverging as $1/n$ in the limit $n \to 0$, a well-defined quantity can be obtained as
\begin{equation}
\lim_{n\to 0} \frac{r^{-2 \Delta_{1}(n)} -r^{-2 \Delta_{2}(n)} }{n} = \kappa \ r^{-2 \Delta_{1}(0)} \log r, 
\label{eqLogMecha}
\end{equation}
where $\kappa= 2 \left. \frac{d(\Delta_{2}-\Delta_{1})}{dn} \right|_{n=0} $ is a universal number. We will say that the two scaling operators corresponding to the critical exponents $\Delta_{1}(n)$ and $\Delta_{2}(n)$ are ``mixed'' at $n=0$, the precise sense of this statement is that the scale transformation operator becomes non-diagonalizable at $n=0$~\cite{Gurarie1993535}. These correlation functions with logarithmic terms can in turn be argued to correspond to logarithmic operators in a LCFT,  so that scale invariance is fully preserved at the critical point despite the presence of non-algebraic correlations. The physical meaning of these logarithmic correlations can then be inferred using Wick's theorem, thus yielding disorder-averaged observables in terms of Green's functions that should behave logarithmically at criticality.

The apparent simplicity of this program is somewhat deceiving: the crucial point is to identify properly the symmetry $G(n)$ of the critical point for generic integer $n$. Note in particular that our approach is drastically different from the classification of operators based on the sigma model formulation of the transition~\cite{Hof:1986aa, Wegner:1987aa, Wegner:1987ab, PhysRevB.87.125144} (see also~\onlinecite{PhysRevLett.112.186803}): after all, the theory describing the critical point probably has nothing to do with Pruisken's sigma model~\cite{Pruisken:1984aa} (or its supersymmetric variants~\cite{Weidenmuller:1987aa}) -- {\it cf.} the above discussion of the WZW model that arises in the strong limit coupling of the $O(3)$ sigma model. On the other hand, the actual symmetry of the critical point is probably much larger that the naive $G(n)$ symmetry identified from the action of the replicated theory. This means that two operators that transform under different irreducible representations of $G(n)$ (and that could eventually be ``mixed'' in the limit $n \to 0$) could actually be part of the same multiplet of fields for the actual, larger symmetry. From a practical point of view, this means that the two exponents $\Delta_1$ and  $\Delta_2$ in eq.~\eqref{eqLogMecha} could be identical, implying that $\kappa=0$, that is, no logarithmic term. Despite this important issue, similar approaches applied to simpler critical points such as the $O(n \to 0)$ model or percolation~\cite{1742-5468-2012-07-L07001,Vasseur2014435} (obtained as the $Q \to 1$ limit of the $Q$-state Potts model with perturbation group $S_Q$ symmetry) were remarkably successful to classify the operator content and the logarithmic correlations of these critical points, even though it is known that the actual symmetry~\cite{Read2007263} of these theories is in fact much larger than $O(n)$ or $S_Q$. Even though quantum Hall transitions are admittedly much more complicated than say the $O(n=0)$ model or percolation,  it is natural to try to extend this approach to critical points that are also much more exciting from a physical perspective. In the following, we will see that such a symmetry-based analysis with the simplest symmetry groups possible yields very sensible results, in agreement with the existing literature on the subject. In the case of the Spin Quantum Hall Effect (SQHE), a close cousin of the Integer Quantum Hall transition where $SU(2)$ spins replace $U(1)$ charges, we predict the existence of logarithmic corrections in some observable that is found to be in agreement with another approach, based on supersymmetry and a mapping onto percolation~\cite{PhysRevLett.82.4524}. For the IQHE transition, we show that the disorder average of a simple combination of Green's functions should scale exactly as $\sim \log r$ at the critical point, without any power-law contribution; a concrete prediction that could be verified numerically using the Chalker-Coddington network model~\cite{0022-3719-21-14-008}.  

We emphasize that extrapolating the limit $n \to 0$ of the replica limit is notoriously complicated in the context of the IQHE -- a fact that can essentially be traced back to the fact that the transition at $\theta=\pi$ is believed to be of first order for $n > 2$, making very hard to extrapolate reliably physical results to $n=0$~\cite{Affleck:1985aa,Affleck:1986aa}. Nevertheless, the replica trick seems more reliable when it comes to predicting the existence of logarithmic correlations (compared to say, computing the exact value of universal quantities like critical exponents in the limit $n \to0$). The reason for this is that the basic mechanism that predicts the emergence of logarithms for $n\to 0$, eq.~\eqref{eqLogMecha}, does not rely on the precise value of the critical exponents or on details of the extrapolation, so that one can essentially ignore that the transition becomes of first order for larger $n$. 
This is well illustrated in the simple case of the $n\to 1$ limit of the $\mathbb{CP}^{n-1}$ sigma model with topological $\theta$-term, which also has a phase transition at $\theta=\pi$. Despite the fact that the transition is also believed to be of first order for $n>2$, we will see below that the replica trick based on the (at least) $U(n)$ symmetry of the critical point leads to results that are in agreement with a more rigorous approaches that are available in this case~\cite{Read2001409,Read2007316,2014arXiv1409.0167G}. 

The remainder of this paper is organized as follows. In section~\ref{SecLoopModel}, we analyze the $n\to 1$ limit of the $\mathbb{CP}^{n-1}$ sigma model, a quantum field theory describing the critical behavior of dense loops with fugacity $n$, a problem apparently unrelated to Anderson transitions. We describe in details how logarithmic correlations arise in the $n\to 1$ limit (corresponding to the classical percolation problem). Building on these results, we analyze the replica limit of the SQHE transition in section~\ref{secSQHE} and predict the existence of logarithmic correlations at the critical point. We show that the same result can also be obtained from the analysis of the percolation problem in Sec.~\ref{SecLoopModel} (see also ref.~\onlinecite{1742-5468-2012-07-L07001}) using the supersymmetry approach developed in Ref.~\onlinecite{PhysRevLett.82.4524}. We then extend the replica approach to the IQHE  case in section~\ref{secIQHE}, and construct an observable that should behave purely logarithmically at the critical point. Finally, section~\ref{secConclusion} contains a discussion of the results and concluding remarks. The main results of our analysis are given by eqs.~\eqref{defOmegaSQHE} and~\eqref{eqLogSQHEOmega} for the SQHE, and eqs.~\eqref{defOmegaIQHE} and~\eqref{eqLogIQHEOmega} for the IQHE -- see also eq.~\eqref{eqLogOnly}.

\section{Warm-up: $n \to 1$ limit of the $\mathbb{CP}^{n-1}$ sigma model, percolation and dense loop models}

\label{SecLoopModel}

We begin with a discussion of logarithmic correlations in a model of dense loops with $SU(n)$ symmetry, a problem apparently completely unrelated to Anderson transitions. However, most results of this paper regarding Anderson transitions rely heavily on this section. The aim of this first section is three-fold: (1) it will provide us with a simple example where the emergence of logarithmic correlations can be understood in details, (2) it introduces the basis of $SU(n)$ representation theory that will be crucial in the following,  and (3) it proves the existence of a logarithmic observable in a critical dense loop model with fugacity $n=1$ (percolation), a model that was shown~\cite{PhysRevLett.82.4524} to describe a certain class of observables in the SQHE transition using the so-called supersymmetry trick. We will show in the next section how this result can be interpreted in the SQHE language using this supersymmetry mapping. 

\subsection{Dense loop models and $SU(n)$ symmetry}

\begin{figure}
\centering
    \includegraphics[width=\columnwidth]
    {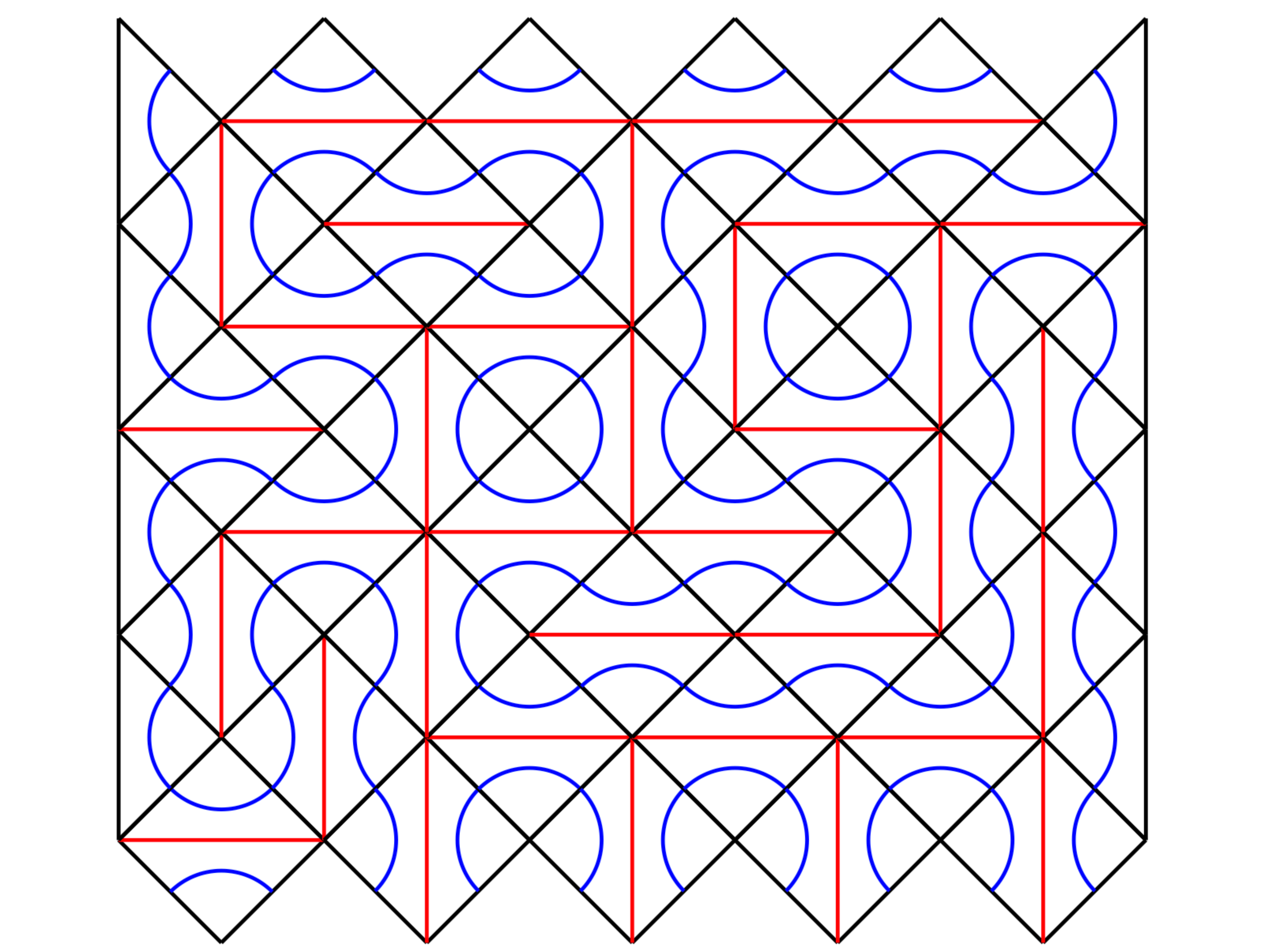}
    \caption{Configuration of a dense loop model with fugacity $n=1$, and corresponding percolation clusters in red. }
\label{FigGeomPercoConfig}
\end{figure}

Let us consider a dense loop model on the square lattice, with fugacity (Boltzmann weight per closed loop) $n \in {\mathbb R}$. An example of configuration of such a loop model is shown in Fig.~\ref{FigGeomPercoConfig}, where the dense loops are drawn in blue. These loops can be thought of as ``hulls'' of percolation clusters shown in red in Fig.~\ref{FigGeomPercoConfig} -- more precisely, the red clusters are percolation clusters only for $n=1$; for generic $n$, they coincide with the so-called Fortuin-Kasteleyn clusters that arise in the high-temperature expansion of the Potts model~\cite{0305-4470-9-3-009}. There are two types of plaquettes in Fig.~\ref{FigGeomPercoConfig}: we will denote the plaquettes corresponding to the propagation of two blue lines along the vertical (imaginary time) direction (with a vertical red line) by ${\bf 1}$, and we will soon see why we use this notation of identity operator. On the other hand, the other type of plaquette with a horizontal red line are denoted by an operator $e$ that contracts two blue lines and creates a new pair of lines in imaginary time. We will see in the following that it is useful to think of these loops as worldlines of bosonic particles that get created and annihilated by the operators $e$ during the imaginary time evolution.   

Labeling by $i$ the columns in Fig.~\ref{FigGeomPercoConfig}, the partition function is computed as follows: if $i$ is even, plaquettes of type ${\bf 1}$ are weighted by $(1-p_A)$ whereas plaquettes associated with the operator $e$ get a weight $p_A$. If $i$ is odd, the same applies with the replacement $p_A \to 1-p_B$. Finally, closed loops carry a Boltzmann weight $n$. The isotropic line corresponds to $p_A=p_B$ and one can show that the system is critical  for $p_A=1-p_B$ and $-2\leq n\leq 2$ (for other values of $n$, there is a first order transition). More formally, we introduce the transfer matrix 
\begin{equation}
T=\prod_{i \ {\rm odd}} ((1-p_A){\bf 1}_i + p_A e_i) \prod_{i \ {\rm even}}(p_B{\bf 1}_i + (1-p_B) e_i), \label{eqT}
\end{equation}
which constructs two rows of the loop model. A full loop configuration such as the one in Fig.~\ref{FigGeomPercoConfig} can then be constructed by successive iterations of this transfer matrix. In the strongly anisotropic limit $p_A \to 0$ with $p_A/(1-p_B)$ fixed, the transfer matrix can be recast as $T \simeq \exp(-\sqrt{p_A (1-p_B)}H)$, with the (1+1)D quantum Hamiltonian
\begin{equation}
H = - \epsilon \sum_{i \ {\rm odd}} e_i - \epsilon^{-1} \sum_{i \ {\rm even}} e_i, \label{eqH}
\end{equation}
with $\epsilon=\sqrt{p_A/(1-p_B)}=1$ at the critical point. 

We now discuss a $SU(n)$ representation of this dense loop model for integer $n$. There are several ways to do this~\cite{0953-8984-2-2-016,Read2001409,PhysRevLett.107.110601}; we follow here Ref.~\onlinecite{Read2001409}.
On each site $i$ (to simplify the notations, we implicitly work on a given imaginary time slice of the system so that  $i$ labels the columns of the network as in eqs~\eqref{eqT} and~\eqref{eqH}), we introduce $n$ bosonic operators $b_i^{a\dagger}$, $b_i^{a}$ with commutation relations $[b_i^{a\dagger},b_j^{b} ] =\delta_{ij} \delta^{ab}$ and $a,b=1,\dots,n$. These operators satisfy the following constraint $\sum_a b_i^{a\dagger} b_i^{a}=1$ (one particle per site), so that the $n$ states $\Ket{a}= b_i^{a\dagger} \Ket{0}$, with $ \Ket{0}$ the Fock space vacuum, form the fundamental (resp. anti fundamental or dual) representation of the Lie algebra su$(n)$ for $i$ even (resp. odd), with the generators acting as $Q^{ab}_i=  b_i^{a\dagger} b_i^{b}$ (resp. $Q^{ab}_i=  -b_i^{b\dagger} b_i^{a}$). The operator $e_i$ is then defined as the projector onto the singlet in the tensor product between fundamental and dual if $i$ is even (dual and fundamental for $i$ odd) $e_i = -\sum_{ab} Q^{ab}_{i}  Q^{ba}_{i+1}$ that has the familiar form of a Heisenberg-like coupling. It is then straightforward to verify that the partition function computed as the trace of powers of the transfer matrix~\eqref{eqT} admits a graphical expansion in terms of non-intersecting dense loops -- the worldlines of the bosons, with the weight of each closed loop given by the trace of the identity operator in the fundamental representation (number of particles flowing around the loop), that is, $n$.  Therefore, the transfer matrix (or the corresponding quantum Hamiltonian) of our loop model can naturally be endowed with a $SU(n)$ symmetry (when $n>1$ is integer). 

The quantum field theory description of these loop models (or of the corresponding $SU(n)$ quantum spin chains) is then provided by a sigma model~\cite{Affleck:1985aa,PhysRevLett.66.1773,Read2001409} on the coset $\mathbb{CP}^{n-1}=U(n)/(U(1)\times U(n-1))$ with topological angle $\theta$. We introduce $n$ bosonic fields $z_a$, $a=1,\dots,n$ subject to the constraint $z^\dagger_a z_a=1$ (implicit summation over repeated indices implied), modulo $U(1)$ phases $z_a \sim e^{i \phi_a} z_a$, so that $z \in \mathbb{CP}^{n-1} \simeq S^{2n-1}/U(1)$. With this field content, the Lagrangian density of the sigma model reads
\begin{equation}
{\cal L}=\frac{1}{2g^2} \left| D_\mu z_a \right|^2 + \frac{i \theta}{2 \pi} \epsilon_{\mu \nu} \partial_\mu a_\nu, 
\end{equation}
where $a_\mu$ is a $U(1)$ gauge field and $D_\mu = \partial_\mu + i a_\mu$. Integrating out the abelian gauge field yields $a_\mu=\frac{i}{2} (z_a^\dagger \partial_\mu z_a - (\partial_\mu z_a^\dagger) z_a)$. The $\theta$-angle contribution is a topological term associated with the non-trivial homotopy group $\pi_2(\mathbb{CP}^{n-1}) = \mathbb{Z}$.

If we now think of $n$ as a real parameter after a naive analytical continuation, it can be shown from the $\beta$ function of the sigma model~\cite{Wegner:1989aa} that the coupling flows to larger values at larger length scales (lower energies) for $n \geq 0$. If $\theta \neq 0$, the system flows to a massive phase with restored $U(n)$ symmetry, but there is a transition at topological angle $\theta=\pi$.  For $n \leq 2$, this transition is believed to be of second order and should therefore be described by a conformal field theory (CFT). Many properties of this CFT as a function of $n$ are known~\cite{Read2001409}, including for example the expression of the central charge and of critical exponents, and in the following we will argue that logarithmic correlations arise naturally from the analytic continuation of the $SU(n)$ symmetry at the value $n=1$ (corresponding to the percolation problem). Note that although we will focus in what follows on this second quantized description of  $SU(n)$ loop models in two dimensions, most of our results also apply to higher dimensional loop models~\cite{PhysRevLett.107.110601,PhysRevB.88.134411}.

\subsection{Observables in  $SU(n)$ loop models}

In order to classify the observables of the loop models, we use the underlying $SU(n)$ symmetry of the critical point. Note that the actual symmetry at the critical point might be much larger than $SU(n)$ -- for these  simple loop models, the symmetry is known~\cite{Read2007263} and is indeed much larger than $SU(n)$ . As a result, observables transforming under different irreducible representations of $SU(n)$ might be part of a larger multiplet -- and in particular would have the same scaling dimension -- under the eventual larger symmetry. We will nevertheless keep working with the $SU(n)$ symmetry and analyze carefully the results in the end. We will focus for concreteness on the two-dimensional second quantized model introduced above, but almost identical results apply to higher dimensional loop models~\cite{PhysRevLett.107.110601,PhysRevB.88.134411}.

\subsubsection{Observables acting on $N=1$ site}

Let us start by classifying the hermitian operators acting on a single edge of the form ${\cal O}(i) = \sum_{a,b} O_{ab} b^\dagger_a(i) b_b(i)$. Clearly, the total number of bosons $\Phi(i)=\sum_c b^\dagger_c(i) b_c(i)$ is invariant under $SU(n)$, and form an irreducible unidimensional representation. Within the context of our loop models, $\Phi(i)=1$ is the identity operator, with scaling dimension $\Delta_\Phi=0$. The remaining $n^2-1$ operators $\phi_{ab}(i)=  b^\dagger_a(i) b_b(i) - \frac{\delta_{ab}}{n}\sum_c b^\dagger_c(i) b_c(i)$ transform irreducibly and form the adjoint representation of $SU(n)$. They are traceless in the sense that they satisfy $\sum_{ab} \delta_{ab} \phi_{ab}(i)=0$. This decomposition can be written in terms of Young diagrams as $[1] \otimes [1]^\star = [1] \otimes [n-1] = [n] \oplus [n-1,1] $. Here and in the sequel, Young diagrams are denoted as $[\lambda_1,\lambda_2,\dots]$, where $\lambda_i$ is the number of boxes in the $i^{\rm th}$ column. Here, $[1]$ is the fundamental (defining) representation of dimension $n$ and $[1]^\star$ its conjugate (dual), and $ [n-1,1] $ is the adjoint. Using the fact that ${\rm tr} \phi_{ab}(i) =0$, where the trace is taken in the fundamental representation of $SU(n)$ generated by $\Ket{a}=b^\dagger_{a} \Ket{0}$, it is then straightforward to show that the two-point function of $\phi_{ab}$ can be expressed as
\begin{equation}
\langle \phi_{aa}(i)  \phi_{bb}(j) \rangle = \frac{1}{n} \left( \delta_{ab} - \frac{1}{n}\right) {\mathbb P}^{(1)}_1(i,j),
\label{eq2leglattice}
\end{equation}
where ${\mathbb P}^{(1)}_1(i,j)$ is the probability than $i$ and $j$ belong to the same loop. The factor $ \left( \delta_{ab} - \frac{1}{n}\right) $ is completely fixed by the condition $\sum_a \phi_{aa}=0$. Note that we have fixed some indices for simplicity but the full correlation function $ \langle \phi_{ab}(i)  \phi_{cd}(j) \rangle  $ can of course be computed similarly. One can also readily check that $\langle \Phi(i) \phi_{ab}(j)\rangle=0 $, and $\langle \phi_{ab} \rangle = \langle \Phi \rangle =0$. 

This exact lattice formula makes the nature of the operator $\phi$ very transparent: it creates a propagating loop, or more precisely, it creates two ``legs'' at site $i$, one incoming and one outgoing. In two dimensions, this observable is known as a ``2-leg watermelon'' operator for this reason. Let $\Delta_\phi(n)$ be its scaling dimension, known exactly in two dimensions~\cite{PhysRevLett.58.2325}: $\Delta_\phi(n) = 1-\frac{2}{g}$, with $g \in [2,4]$ given by $n^2=2(1+\cos \frac{\pi g}{2})$. In the continuum limit, we expect the following scaling
\begin{equation}
\langle \phi_{aa}(r_i)  \phi_{bb}(r_j) \rangle = A(n) \left( \delta_{ab} - \frac{1}{n}\right) r^{-2 \Delta_\phi(n)},
\label{eq2legcontinuum}
\end{equation}
where $A(n)$ is some non-zero function of $n$, and $r=| r_i - r_j|$.

To summarize, the $SU(n)$ symmetry allowed us to identify two distinct scaling operators, the identity and a 2-leg watermelon operator. While the representation theory analysis was performed for $n$ integer, the correlation functions~\eqref{eq2leglattice},~\eqref{eq2legcontinuum} make sense for generic $n$. Although nothing particularly exciting happens in the limit $n \to 1$, one can already see that the limit $n \to 0$ will be singular from eq.~\eqref{eq2legcontinuum}. As we will see more explicitly in the following, poles like the one in eq.~\eqref{eq2legcontinuum} lead to logarithmic correlations. In this case, one can check that the scaling dimension $\Delta_\phi(n)$ vanishes as $n \to 0$, indicating a ``mixing'' of the operator $\phi_{ab}$ with $\Phi$ (the identity) at $n=0$. This mixing can also be traced back to the dimension $n^2-1$ of the representation of $\phi$ that becomes formally $-1$ in the limit $n\to 0$. We will not describe this limit $n \to 0$ here, but instead turn to more complicated operators that will have singular limits as $n \to 1$. We will then show more explicitly how these poles lead to logarithmic correlations.

\begin{figure}
\centering
   \includegraphics[width=\columnwidth]{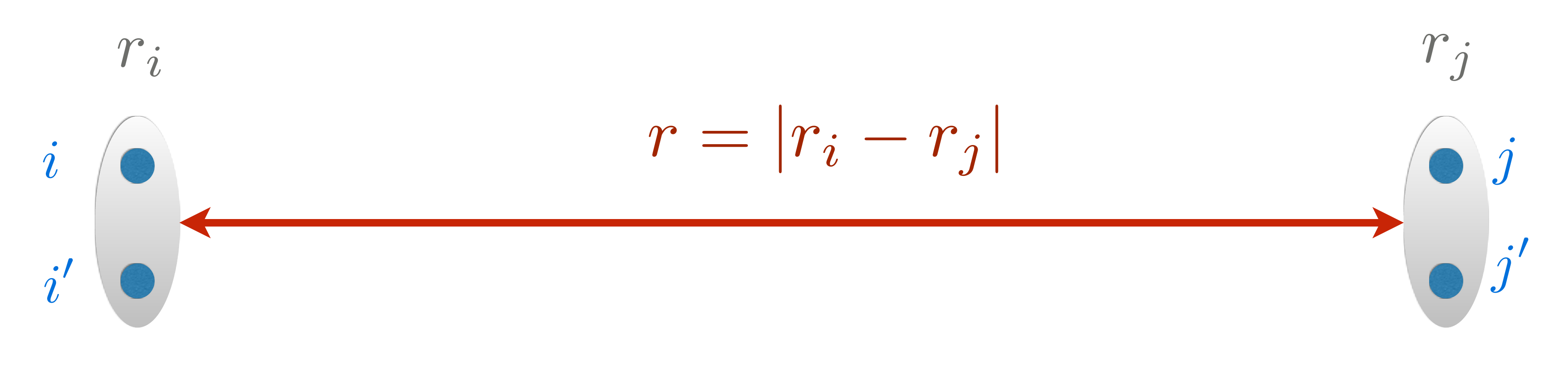}
    \caption{Schematic representation of the coarse grained two-point function $\langle {\cal O}(r_i) {\cal O}(r_j)\rangle $ of observables acting on $N=2$ sites. The two sites $i$ and $i^\prime$ ({\it resp.} $j$ and $j^\prime$) are chosen in the infinitesimal neighborhood of the position $r_i$ ({\it resp.} $r_j$).}
\label{Fig2pointObs}
\end{figure}

\subsubsection{Observables acting on $N=2$ sites}

We now turn to observables acting on two nearest neighbor edges $i$ and $i^\prime$. For simplicity, we consider a coarse-grained picture in which $i$ and $i^\prime$ are two sites near position $r_i$ (see Fig.~\ref{Fig2pointObs}), corresponding to the fundamental representation of $SU(n)$ (even columns). We wish to consider operators of the type $b_a^\dagger(i)b_b(i)b_c^\dagger(i^\prime)b_d(i^\prime)$. It is natural to enforce a symmetry between $i$ and $i^\prime$, and we further restrict our study to operators that are symmetric under the exchange $(a,b) \leftrightarrow (c,d)$, and that vanish if $a=c$ or $b=d$. The last two conditions are conveniently satisfied by operators that are antisymmetric under the exchanges $a \leftrightarrow c$ or $b \leftrightarrow d$. We therefore define $T_{abcd}(r_i)=-Q_{ab}(i)Q_{cd}(i^\prime)+Q_{ad}(i)Q_{cb}(i^\prime)-Q_{ab}(i^\prime)Q_{cd}(i)+Q_{ad}(i^\prime)Q_{cb}(i)$, with $Q_{ab}(i)=b^\dagger_a (i)b_b(i)$. These operators transform under $SU(n)$ as $[2] \otimes [2]^\star \subset ([1] \otimes [1]^\star)^{\otimes 2}$, with $(n(n-1)/2)^2$ components. This representation is reducible, and can be decomposed as $[2] \otimes [2]^\star = [n] \oplus [n-1,1] \oplus [n-2,2]$. 

The first term is the invariant $T^0=\sum_{ab} T_{aabb}$ where all the indices are contracted. Defining ${\rm Tr} Q = \sum_a Q_{aa}$ -- not to be confused with the symbol ${\rm tr}$ which corresponds to the trace in the fundamental representation -- this operator can be expressed as $T^0(r_i) = - 2{\rm Tr } Q(i) {\rm Tr} Q(i^\prime) + 2{\rm Tr} Q(i) Q(i^\prime) =2({\rm Tr} Q(i) Q(i^\prime) -1) $, which is clearly invariant. This operator corresponds to a lattice version of the energy operator. More precisely, we define the energy operator $\varepsilon(r_i)=T^0(r_i) - \langle T^0 \rangle$, where we have subtracted the constant term $\langle T^0 \rangle = 2(n-1) {\mathbb P}^{(1)}_1(i, i^\prime)+2(1/n-1) {\mathbb P}^{(1)}_0(i, i^\prime)$, where ${\mathbb P}^{(1)}_0(i, i^\prime)$ is the probability that $i$ and $i^\prime$ belong to different loops, {\it i.e.} ${\mathbb P}^{(1)}_0(i, i^\prime) = 1- {\mathbb P}^{(1)}_1(i, i^\prime)$. Note that since the sites $i$ and $i^\prime$ lie within a small neighborhood around the position $r_i$, $\langle T^0 \rangle$ is indeed a constant in a coarse grained picture, {\it i.e.}, it does not depend on $r_i$. The two point function $\langle\varepsilon(r_i) \varepsilon(r_j) \rangle$ can be expressed exactly in terms of lattice probabilities conditioning  the points $i$, $i^\prime$, $j$ and $j^\prime$ to belong to  various loops. The explicit result is not very illuminating, but it shows that $\langle\varepsilon(r_i) \varepsilon(r_j) \rangle$ vanishes as $n\to 1$. Denoting by $\Delta_{\varepsilon}(n)$ the scaling dimension of $\varepsilon(r_i)$, we thus expect, in the continuum limit 
\begin{equation}
\langle \varepsilon(r_i)  \varepsilon(r_j) \rangle = A_0(n) (n-1) r^{-2 \Delta_\varepsilon(n)},
\label{eqEnergycontinuum}
\end{equation}
where $A_0(n)$ is a regular function of $n$ with $A_0(1)\neq 0$ finite, and $r=| r_i - r_j|$. In two dimensions, we have~\cite{PhysRevB.23.429} $\Delta_{\varepsilon}(n)=\frac{6}{g}-1$, where $g$ was defined above.

\begin{figure}
\centering
    \includegraphics[width=\columnwidth]
    {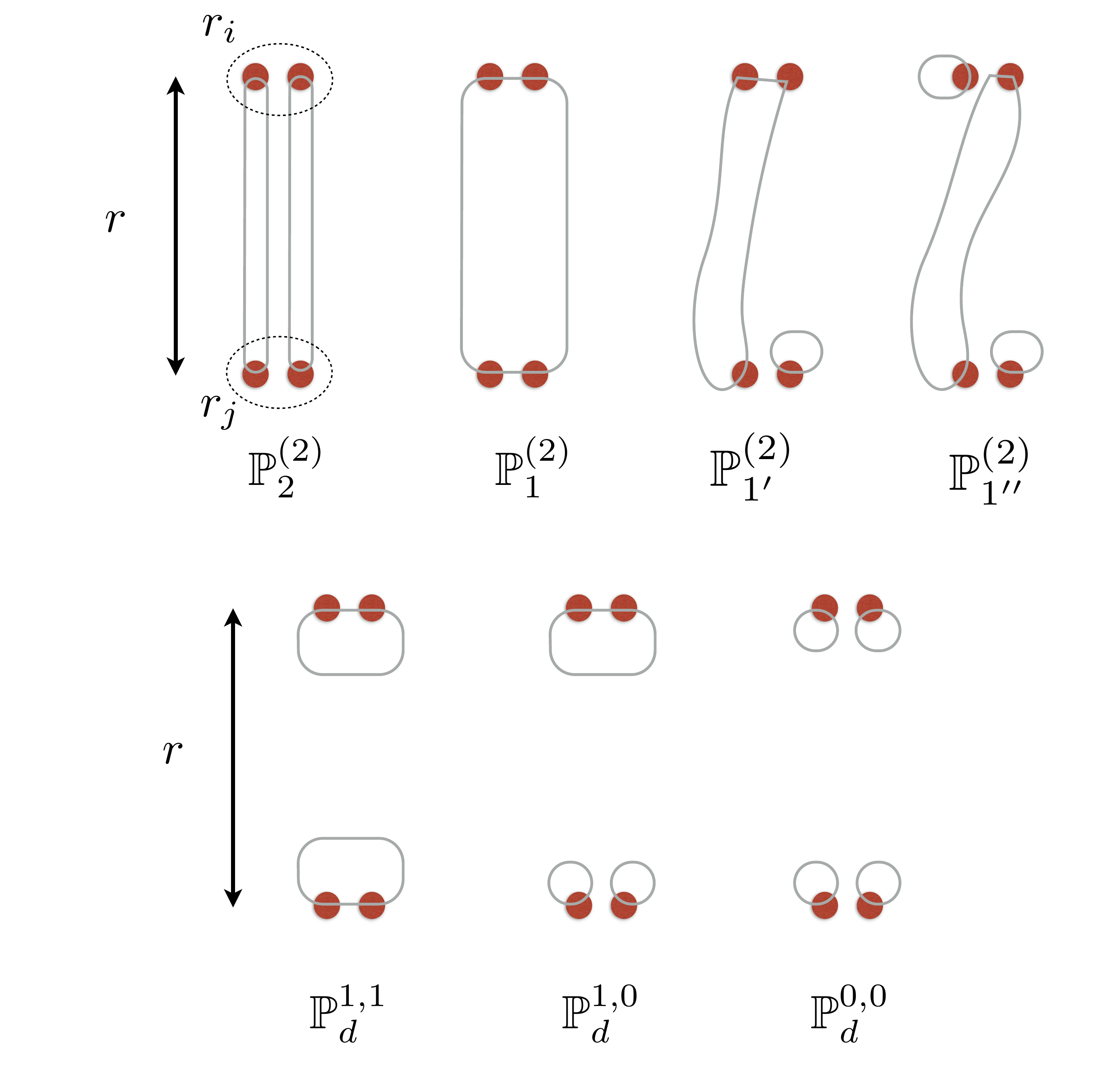}
    \caption{Graphical representations of loop configurations contributing to the various probabilities introduced in the text. Note that some of the probabilities contain more terms than the ones drawn: for instance, ${\mathbb P}^{(2)}_{1^\prime}$ counts configurations in which any of the four points is isolated in a loop while the other three points belong to another loop. See main text for a precise definition of all these probabilities. }
\label{FigGeomPerco}
\end{figure}

The second term $[n-1,1] $ in the decomposition of $[2] \otimes [2]^\star$ is the adjoint representation, corresponding to the $n^2-1$ fields $T_{ab}^1(r_i)=\sum_{c} T_{abcc} -\frac{\delta_{ab}}{n} T^0= - Q(i) {\rm Tr} Q(i^\prime) + Q(i)Q(i^\prime) + i \leftrightarrow i^\prime -\frac{\delta_{ab}}{n} T^0$. This operator has the same symmetry as $\phi_{ab}(r_i)$, we thus expect its two-point function to be dominated by a power-law behavior with the same scaling dimension $ \Delta_\phi(n)$, with some eventual subleading contributions. This correlator can also be expressed on the lattice as 
\begin{multline}
\langle T^1_{aa}(r_i)  T^1_{bb}(r_j) \rangle =\frac{1}{n} \left(\delta_{ab}-\frac{1}{n}\right)\frac{(n-2)}{n}\times \left[ 4 {\mathbb P}^{(2)}_2(r_i,r_j) \right. \\
\left. + (n-2)(n {\mathbb P}^{(2)}_1(r_i,r_j)-{\mathbb P}^{(2)}_{1^\prime}(r_i,r_j)+\frac{1}{n}{\mathbb P}^{(2)}_{1^{\prime\prime}}(r_i,r_j))\right].
\label{eqT1Lattice}
\end{multline}
In this expression, ${\mathbb P}^{(2)}_2(r_i,r_j) $ is the probability that $i$ and $j$ belong to the same loop while $i^\prime$ and $j^\prime$ belong to another loop, or that $i$ and $j^\prime$ belong to the same loop while $i^\prime$ and $j$ belong to another loop. ${\mathbb P}^{(2)}_1(r_i,r_j)$ is the probability that the four points $i$, $i^\prime$, $j$ and $j^\prime$ belong to the same loop, whereas ${\mathbb P}^{(2)}_{1^\prime}(r_i,r_j)$ counts configurations in which three out of the four points belong to the same loop, while the last remaining point in its own loop (see Fig.~\ref{FigGeomPerco}). Finally,  ${\mathbb P}^{(2)}_{1^{\prime \prime}}(r_i,r_j)$ corresponds to configurations where either $i$ or $i^\prime$ is in the same loop than either $j$ or $j^\prime$, with the remaining two points being alone in their own loop.  All these probabilities vanish as $r=|r_i -r_j| \to \infty$.

Finally, the last term $[n-2,2]$ in $[2] \otimes [2]^\star$ has dimension $n^2(n+1)(n-3)/4$, which becomes formally $-1$ when $n=1$. As we will describe in details in the following, this is the representation that will be mixed with the energy operator when $n=1$. These operators are given explicitly by $T^{2}_{abcd}=T_{abcd}-\frac{1}{n-2}\left(\delta_{ab} T^1_{cd}+\delta_{cd} T^1_{ab}- \delta_{bc} T^1_{ad} -\delta_{ad} T^1_{cb} \right)-\frac{1}{n(n-1)} \left(\delta_{ab}\delta_{cd} -\delta_{ad}\delta_{cb} \right)T^0$, and they satisfy the $n^2$ constraints $\sum_a T^2_{aacd}=0$. The two point function of $T^2$ then reads
\begin{multline}
\langle T^2_{aabb}(r_i)  T^2_{ccdd}(r_j) \rangle = \frac{4}{n^2}\left(\delta_{ac}\delta_{bd}+\delta_{ad}\delta_{bc}-\frac{1}{n-2} \times \right. \\ \left.(\delta_{ac}+\delta_{bd}+\delta_{ad}+\delta_{bc})+ \frac{2}{(n-1)(n-2)} \right){\mathbb P}^{(2)}_{2}(r_i,r_j),
\label{eqT2Lattice}
\end{multline}
with $a \neq b$ and $c \neq d$ -- otherwise this correlator is simply zero.
Despite this rather complicated expression, this correlator is remarkably simple. The combination of Kronecker deltas is entirely fixed by the symmetry constraints $\sum_{a=c,d} T^2_{aacd}=0$, and the dependence on ${\mathbb P}^{(2)}_{2}(r_i,r_j)$ provides a natural physical interpretation of $T^2$ as a ``4-leg watermelon operator'', with two loops propagating from $r_i$ to $r_j$. One can also check that all the crossed two-point correlation functions between $T^2$, $T^1$ and $\varepsilon$ vanish, as required by symmetry. The scaling dimension of $T^2$ reads~\cite{PhysRevLett.58.2325} $\Delta_{T^2} (n)=\frac{(4+g)(3g-4)}{8g}$, and in the scaling limit, we expect
\begin{multline}
\langle T^2_{aabb}(r_i)  T^2_{ccdd}(r_j) \rangle = \frac{2 A_2(n)}{n^2}\left(\delta_{ac}\delta_{bd}+\delta_{ad}\delta_{bc}-\frac{1}{n-2} \times \right. \\ \left.(\delta_{ac}+\delta_{bd}+\delta_{ad}+\delta_{bc})+ \frac{2}{(n-1)(n-2)} \right) r^{-2 \Delta_{T^2} (n)}.
\label{eqT2continuum}
\end{multline}

\subsubsection{Other observables}

It is obviously possible to generalize the above to observables acting on $N>2$ sites, but this will not be necessary for our purposes. Moreover, it would be natural to ask what happens to the symmetric observables $\tilde{T}_{abcd}(r_i)=Q_{ab}(i)Q_{cd}(i^\prime)+Q_{ad}(i)Q_{cb}(i^\prime)+Q_{ab}(i^\prime)Q_{cd}(i)+Q_{ad}(i^\prime)Q_{cb}(i)$, which would also transform reducibly under $SU(n)$. We chose to focus on the antisymmetric tensors $T_{abcd}(r_i)=-Q_{ab}(i)Q_{cd}(i^\prime)+Q_{ad}(i)Q_{cb}(i^\prime)-Q_{ab}(i^\prime)Q_{cd}(i)+Q_{ad}(i^\prime)Q_{cb}(i)$ to forbid the cases $a=c$ or $b=d$, which would effectively reduce to operators acting on $N=1$ site in a coarse grained picture. It is of course possible to carry on the same analysis for $\tilde{T}_{abcd}$ instead of $T_{abcd}$, and one ends up with three irreducible representations with dimensions 1, $n^2-1$ and $n^2(n+3)(n-1)/4$ -- note that the dimension of this last representation becomes formally $0$ at $n=1$ (instead of $-1$ in the antisymmetric case above), indicating the absence of mixing. One finds the same pole structure as for $N=1$, and in particular, the limit $n\to 1$ of all the irreducible operators constructed out of $\tilde{T}_{abcd}$ is well defined. In the end, we find that the correlation functions of these other operators in the limit $n \to 1$ do not contain any new information so that it is possible to focus only on the observables considered in the previous paragraph. 

\subsection{Logarithmic correlations in the $n \to 1$ limit}
\label{subseclogPerco}

The $n \to 1$ of eq.~\eqref{eqT2continuum} is clearly ill-defined. This indicates a ``mixing'' between the scaling operators $T^2$ and $\varepsilon$ at $n=1$, and is further confirmed by the fact that the scaling dimensions $\Delta_\varepsilon=\Delta_{T^2}=\frac{5}{4}$ coincide for that value of $n$.
In order to obtain finite correlation functions for $n \to 1$, we follow~\cite{1999cond.mat.11024C} (see also~\cite{1742-5468-2012-07-L07001}) and introduce a new field $\psi_{ab} = (1-\delta_{ab})(T^2_{aabb}+\frac{1}{n(n-1)} \varepsilon)$. Finiteness of correlation functions at $n=1$ then requires $A_0(1)=A_2(1)$, and one ends up with the logarithmic correlation function at $n=1$
\begin{multline}
\langle \psi_{ab}(r_i)  \psi_{cd}(r_j) \rangle = 2 A_0(1) r^{-5/2} \left( \delta_{ac}\delta_{bd}+\delta_{ad}\delta_{bc}
\right. \\ \left.+\delta_{ac}+\delta_{bd}+\delta_{ad}+\delta_{bc} +\kappa \log \frac{r}{a} \right),
\label{eqLogLoop}
\end{multline}
where $a$ is the lattice spacing (UV cutoff), and $\kappa$ is universal and given by
\begin{equation}
\kappa= 4 \lim_{n \to 1} \frac{\Delta_{T^2}-\Delta_{\varepsilon}}{n-1} = \frac{8 \sqrt{3}}{\pi}.
\label{eqDelta}
\end{equation}
This proves the emergence of a field with logarithmic correlations at the critical point. Moreover, one can readily check that $\psi_{ab}$ is mixed with the energy operator under a scale transformation $r \mapsto \Lambda r$, a feature characteristic of logarithmic operators~\cite{Gurarie1993535}: $\psi_{ab}(\Lambda r) = \Lambda^{-5/4} \left( \psi_{ab}(r) + \frac{\kappa}{2} \log\Lambda \ \epsilon(r)\right)$, whereas  $\epsilon(\Lambda r) = \Lambda^{-5/4} \epsilon(r)$. We also note that the Kronecker delta in~\eqref{eqLogLoop} are purely formal since $n=1$, but it will turn out to be convenient to keep track of this structure to interpret this correlation function geometrically. 

In order to elucidate the geometrical meaning of this logarithmic two-point function, it is fruitful to go back to lattice correlation functions. Using eq.~\eqref{eqT2Lattice} and the exact expression of the two-point function of $\varepsilon$ in terms of probabilities, one can readily express the two-point function of $\psi_{ab}$ in terms of various probabilities as well. The now well-defined $n \to 1$ limit then yields 
\begin{multline}
\langle \psi_{ab}(r_i)  \psi_{cd}(r_j) \rangle = 4 {\mathbb P}^{(2)}_{2}(r) \left( \delta_{ac}\delta_{bd}+\delta_{ad}\delta_{bc}\right. \\ \left.+\delta_{ac}+\delta_{bd}
+\delta_{ad}+\delta_{bc}  \right)+ 4 \left( F_d(r)-  F_d^\infty +F_c(r)\right).
\label{eqLogLoopLattice}
\end{multline}
In this expression, $F_d(r_i,r_j) = {\mathbb P}^{1,1}_d+{\mathbb P}^{0,0}_d-{\mathbb P}^{1,0}_d$ is a linear combination of {\it disconnected} probabilities (Fig.~\ref{FigGeomPerco}): ${\mathbb P}_d^{0,0}$ is the probability that the four points $i$, $i^\prime$, $j$ and $j^\prime$ belong to four different loops, ${\mathbb P}_d^{1,1}$ is the probability that $i$ and $i^\prime$ belong to the same loop while $j$ and $j^\prime$ belong to another loop, ${\mathbb P}_d^{1,0}$ is the probability that $j$ and $j^\prime$ belong to the same loop while $i$ and $i^\prime$ belong to two other loops, or the same with the role of $r_i$ and $r_j$ reversed. $F_d(r_i,r_j) $ has a non-vanishing limit when $r=|r_i-r_j| \to \infty$ given by $F_d^\infty = ({\mathbb P}^{(1)}_1(i,i^\prime)- {\mathbb P}^{(0)}_1(i,i^\prime))^2=(1-2 {\mathbb P}^{(0)}_1(i,i^\prime))^2 $, which is a constant independent of the position $r_i$ -- it is however non-universal and depends on how the coarse graining procedure is defined. Finally, $F_c(r_i,r_j)$ corresponds to the connected part and is given by $F_c(r_i,r_j)={\mathbb P}^{(2)}_1(r_i,r_j)-{\mathbb P}^{(2)}_{1^\prime}(r_i,r_j)+{\mathbb P}^{(2)}_{1^{\prime \prime}}(r_i,r_j)-4  {\mathbb P}^{(2)}_{2}(r)$.

Comparing eq~\eqref{eqLogLoop} and eq~\eqref{eqLogLoopLattice}, we immediately find $2 {\mathbb P}^{(2)}_{2}(r)  \sim A_0(1) r^{-5/2}$ and $ 2(F_d(r)-  F_d^\infty +F_c(r))  \sim A_0(1) r^{-5/2} \kappa \log \frac{r}{a}$. In particular, the ratio 
\begin{equation}
\frac{F_d(r)-  F_d^\infty +F_c(r)}{{\mathbb P}^{(2)}_{2}(r)} \sim \kappa \log \frac{r}{a},
\label{eqLogPerco}
\end{equation}
should be purely logarithmic for a dense loop model with fugacity $n=1$, with the universal amplitude $\kappa$ given by~\eqref{eqDelta}. We expect this expression to be particularly suited for numerical checks, since it isolates the logarithmic dependence. It is worth pointing out that the logarithmic dependence stems only from the disconnected part $F_d(r)$: using the $n\to 1$ limit of~\eqref{eqT1Lattice}, one can easily show that the connected piece $F_c(r) \propto \langle T^1_{aa}(r)  T^1_{bb}(0) \rangle$ is purely algebraic for $n \to 1$, with a  leading contribution given by the 2-leg exponent $\Delta_{\phi}(n=1)=\frac{1}{4}$.

\subsection{Relation to the Potts model, percolation, and supersymmetry}

\subsubsection{Percolation and Potts model}
We have thus shown that dense loop models with fugacity $n=1$ involve a logarithmic operator resulting from the mixing of the energy and the 4-leg watermelon fields. The two point function of this logarithmic operator can be expressed in terms of probabilities that could readily be measured using Monte Carlo simulations for example. When $n=1$, the two dimensional dense loop model introduced above can be mapped onto bond percolation, where the loops are the hulls of percolation clusters. Our results thus also apply to the two-dimensional percolation problem, where the observables that we described act on the hulls of percolation clusters. The existence of a logarithmic operator mixing energy and 4-leg fields in percolation was pointed out in Refs.~\onlinecite{PhysRevLett.108.161602,1742-5468-2012-07-L07001}, and a similar concrete logarithmic observable in percolation was uncovered~\cite{1742-5468-2012-07-L07001,Vasseur2014435} using the $Q \to 1$ limit of the Potts model. Despite the similarities between the above calculation and the result of Ref.~\onlinecite{1742-5468-2012-07-L07001} -- that can be traced back to the relation between the representation theory of $SU(n)$ and the symmetric group -- note that the resulting logarithmic observables are different, since the ones obtained from the Potts model act on percolation clusters, instead of acting on hulls. Therefore, even if the Potts model is known to be related to loop models with $SU(n)$ symmetry~\cite{0953-8984-2-2-016}, the observables in both models are different. For instance, the probability that two points separated by a distance $r $ belong to the same percolation cluster is definitely different from the probability that two points belong to the same loop surrounding a cluster, and they actually scale with different exponents, $r^{-5/24}$ and  $r^{-1/2}$ respectively~\cite{PhysRevB.27.1674,PhysRevLett.58.2325}.  

\subsubsection{Supersymmetry vs Replicas}
\label{secSUSYperco}
Although the loop models introduced above do not involve quenched disorder, it is helpful to think of our approach based on the formal limit $n\to 1$ of the $SU(n)$ symmetry as a sort of replica trick. As in disordered systems, it is possible to use an alternative, more rigorous approach based on supersymmetry to construct loop models~\cite{PhysRevLett.82.4524,Read2001409,PhysRevLett.90.090601}. Intuitively, one introduces both fermonic and bosonic operators to ensure that closed loops get a weight $n=1$, the fermonic terms giving negative contributions when computing the weight. Loop models with fugacity $n$ are then described by a $SU(n+m|m)$ global (super)symmetry (SUSY), where $SU(n+m|m)$ is the supergroup analog to $SU(n)$, defined by the transformations preserving a form with $n+m$ bosonic variables and $m$ fermonic ones (see {\it e.g.} Ref.~\onlinecite{1996hep.th7161F} for a review of Lie superalgebras). There is then no issue with the value $n=1$ and percolation can be described~\cite{PhysRevLett.82.4524,Read2001409} by, {\it e.g.} a theory with $SU(2|1)$ SUSY (or $SU(m+1|m)$ SUSY in general). The emergence of logarithmic correlations can also be understood in this supersymmetric language~\cite{0305-4470-35-27-101,2004hep.th9105G, Read2007316}, using more involved representation theory tools, such as indecomposable representations -- representations that are reducible but not fully reducible. We will not describe the SUSY approach in more detail here (see~\onlinecite{Read2001409}), but we simply note that although SUSY is very powerful to predict formally the emergence of logarithmic operators, it does not yield simple geometrical formulas like~\eqref{eqLogPerco}.

\subsubsection{Extended symmetry}
\label{subsubsecExtended}

As mentioned above, the actual symmetry at the strong coupling fixed point of the two-dimensional $\mathbb{CP}^{n-1}$ sigma model is much larger than $SU(n)$~\cite{Read2007263}. As a result, operators that transform under different irreducible representations of $SU(n)$ could actually be part of the same multiplet of fields for the actual, larger symmetry. This means that the prefactors of some of the logarithms predicted using the $SU(n)$ symmetry could in fact be zero. In the example~\eqref{eqLogLoop} described above, the prefactor~\eqref{eqDelta} is non-zero and can be computed exactly. Moreover, the true symmetry of the critical point is also known exactly for those dense loop models~\cite{Read2007263}. For example, the representation $[n-2,2]$ with dimension $n^2(n+1)(n-3)/4$ of $SU(n)$ is part of a much larger representation of dimension $n^2(n^2-3)/2$ for the actual symmetry~\cite{Read2007263} (or $n^4-3 n^2 +1$ for fields living at the boundary). However, in the limit $n \to 1$, the dimension of both representations goes (formally) to $-1$, indicating a ``mixing'' (in the sense described above) with the trivial representation of dimension $1$. Therefore, in the limit $n \to1$, the $SU(n)$ symmetry is enough to understand the mixing of the 4-leg operator (corresponding to the representation $[n-2,2]$) and the identity operator. 

Even if in this case, considering a smaller symmetry does not affect the prediction of logarithmic correlations for the 4-leg operator, this example nicely illustrates that fields having different symmetry properties under $SU(n)$ can actually have the same scaling dimension. More precisely, for bulk fields we find that the large representation $n^2(n^2-3)/2$ for the true symmetry decomposes as $(n^2(n+3)(n-1)/4) \oplus (n^2(n-3)(n+1)/4)$ under $SU(n)$, where we denoted representations by their dimension for simplicity.  For boundary fields, the representation $n^4-3 n^2 +1$  for the true symmetry decomposes as $(n^2(n+3)(n-1)/4) \oplus (n^2(n-3)(n+1)/4) \oplus (n^2-1) \oplus 2 \times ((n^2-4)(n^2-1)/4)$. These irreducible $SU(n)$ representations -- all contained in the tensor product $ ([1] \otimes [1]^\star)^{\otimes 2}$ -- therefore correspond to the same primary field (the 4-leg operator) for the extended symmetry. Interestingly, some aspects of that analysis can be recovered directly at the level of the $\mathbb{CP}^{n-1}$ sigma model\footnote{A. Nahum: private communication. }.

Note also that considering a smaller symmetry group at the critical point may also lead us to miss some logarithms: for example, considering the full symmetry of dense loop models predicts a mixing between the representations of dimension $n^2(n^2-3)/2$ (or $n^4-3 n^2 +1$ at the boundary) and $n^2-1$ (adjoint) for $n=\sqrt{2}$, consistent with other approaches~\cite{1742-5468-2012-07-L07001}. However, we would completely miss this mixing by considering only the smaller $SU(n)$ symmetry -- there is no pole at  $n=\sqrt{2}$ in eq.~\eqref{eqT2continuum}. Having these drawbacks in mind, we are now ready to go back to Anderson transitions. 

\section{Logarithmic observables at the Spin Quantum Hall transition}

\label{secSQHE}

We now turn to the study of logarithmic correlations at the Spin Quantum Hall transition, a close cousin of the Integer Quantum Hall transition where $SU(2)$ spin is transported instead of $U(1)$ charge. We first review the basic physics of the Spin Quantum Hall Effect and introduce the corresponding network model that describes the transition. Using the replica trick and the global symmetries at the critical point, we then exhibit linear combinations of Green's functions whose disorder average should be logarithmic. Finally, using supersymmetry and an exact mapping onto classical percolation~\cite{PhysRevLett.82.4524}, we argue that this logarithmic observable found using the replica trick coincides with the logarithmic two-point correlator in the $SU(n=1)$ loop model discussed in the previous section.

\subsection{Spin Quantum Hall Transition}

The Spin Quantum Hall Effect (SQHE) is the quantization of the spin Hall conductance, analogously to the quantization of the Hall charge conductance in the Integer Quantum Hall Effect (IQHE). Spin transport is defined as the response to the spatial variation of an external Zeeman magnetic field, which plays the role of the electric field in the IQHE. If the external field is along the $z$ direction, the spin Hall conductance $\sigma_{xy}^{\rm S}$ is defined by the spin current along the $x$ direction
\begin{equation}
j^z_x = - \sigma_{xy}^{\rm S} \partial_y B_z.
\end{equation}
As in the IQHE, the spin Hall conductance $\sigma_{xy}^{\rm S}$ can take non-zero values only in systems with broken time-reversal invariance. 

The SQHE may occur is singlet superconductors~\cite{PhysRevLett.82.3516, PhysRevB.60.4245} with mean-field Hamiltonian
\begin{equation}
H = \sum_{ij} \sum_{\sigma} t_{ij} c^\dagger_{i \sigma} c_{j\sigma} + \sum_{ij} \Delta_{ij} c^\dagger_{i\uparrow} c^\dagger_{j\downarrow} + \Delta^\star_{ij} c_{j\downarrow} c_{\uparrow}.
\end{equation}
There is no notion of charge transport since charge is obviously not conserved by the superconductor, but  the Hamiltonian has a global $SU(2)$ spin rotation symmetry if $\Delta_{ij}=\Delta_{ji}$. If the gap function $\Delta$ is complex, time reversal invariance is broken and $\sigma_{xy}^{\rm S}$ may take non-zero values. For a clean $d_{x^2-y^2}+i d_{xy}$ superconductor, one can show that $\sigma_{xy}^{\rm S}$ is quantized and that transport is carried by chiral edge modes at the edge of a system with a boundary. These edge states are robust to weak disorder, whereas strong disorder eventually drives the system to a topologically distinct localized phase characterized by $\sigma_{xy}^{\rm S} = 0$ (spin insulator). In the following, we will study this Anderson transition from the SQHE phase to the spin insulating phase.  

Spin rotation symmetry and broken time reversal invariance defines the symmetry class $C$~\cite{10.1063/1.531675,PhysRevB.55.1142}. The corresponding sigma model describing the transition~\cite{PhysRevLett.81.4704,PhysRevB.60.6893} is defined on the coset $Sp(2n)/U(n)$ with $n\to0$ using a replica approach, or $OSp(2|2)/U(1|1)$ using supersymmetry (more general cosets $OSp(2m|2m)/U(m|m)$ can also be used to describe fluctuations of observables and compute multi-fractal exponents). 

\begin{figure}
\centering
    \includegraphics[width=\columnwidth]
    {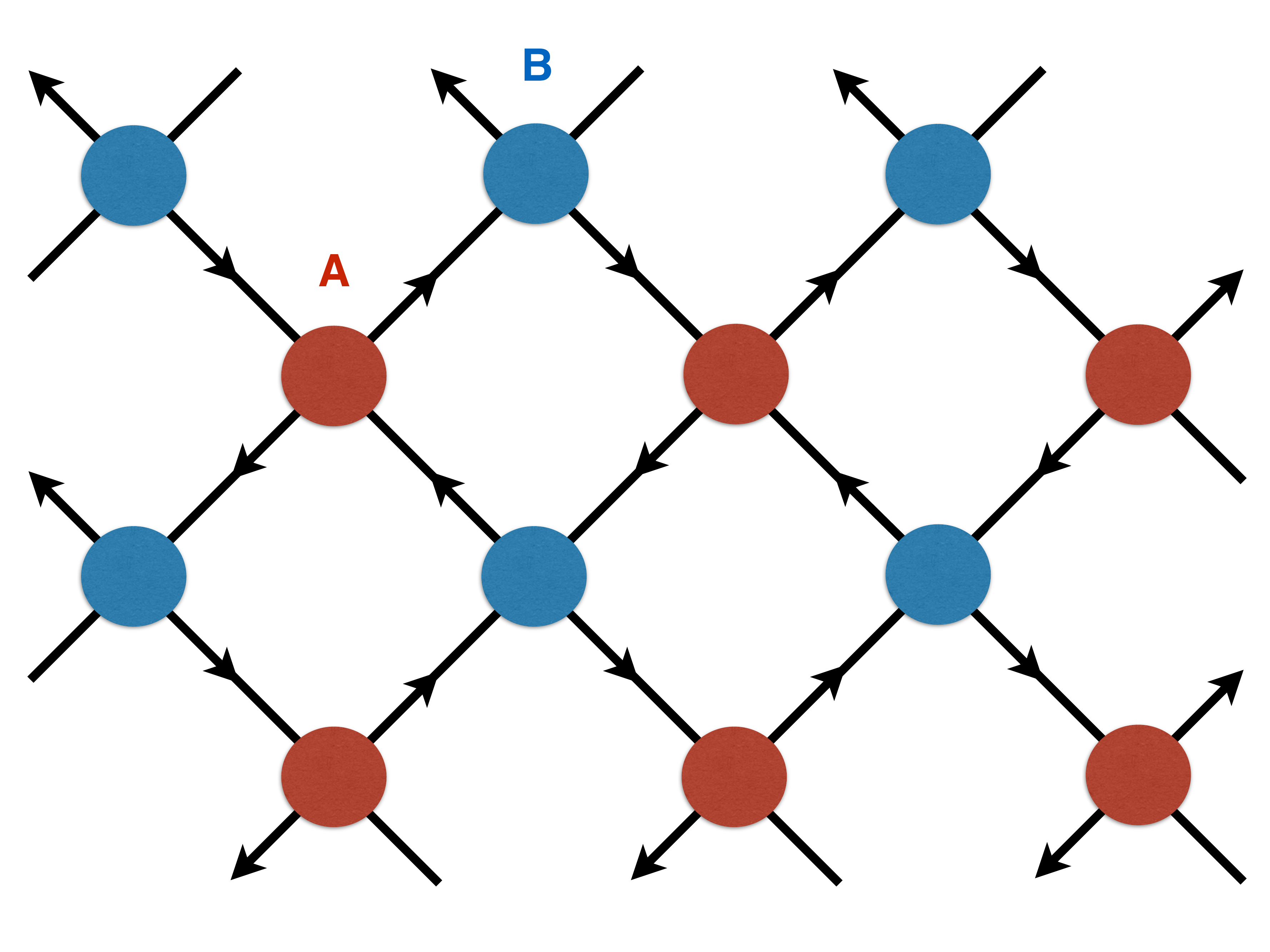}
    \caption{Network model with directed links and two types of nodes A and B on the square lattice. Particles can propagate along the directed links, with random $SU(2)$ (resp. $U(1)$) matrices associated with the propagation on each link in the SQHE (resp. IQHE) case. The blue and red circles represent scattering matrices at each node. }
\label{FigCCNetwork}
\end{figure}

\subsection{Network model and Green's functions}

The lattice version of this non-interacting disordered fermions problem is introduced using a variant~\cite{PhysRevLett.82.3516, PhysRevLett.82.4524} of the Chalker-Coddington network model for the Integer Quantum Hall transition~\cite{0022-3719-21-14-008}. 

As in the  Chalker-Coddington model, we introduce a network with directed links and two types of nodes A and B on the square lattice (Fig.~\ref{FigCCNetwork}) where particles can propagate. The difference with the Chalker-Coddington case comes from the spin of these particles, as disorder is introduced by random $SU(2)$ matrices (instead of random $U(1)$ phases) associated with the propagation along each link. The nodes of the network correspond to scattering events and we take the scattering matrices at each node to be diagonal in the spin subspace and 
\begin{equation}
S_{K,\sigma} =  \left( \begin{array}{cc}
\sqrt{1-t_{K\sigma}^2} &  t_{K\sigma}\\
-t_{K\sigma} & \sqrt{1-t_{K\sigma}^2}  \end{array} \right),
\end{equation}
where $\sigma=\uparrow,\downarrow$ labels the spin of the particles and $K=A,B$ denotes the two different types of nodes. Spin rotation symmetry implies that $t_K \equiv t_{K\uparrow}=t_{K\downarrow}$\footnote{Taking $t_{K\uparrow}\neq t_{K\downarrow}$ breaks the $SU(2)$ symmetry and splits the transition into two copies of the IQHE.  }, and isotropy of the network enforces $t_{A}^2+t_B^2=1$, so that we parametrize $t_A=\cos \theta$ and $t_B=\sin \theta$. The isotropic critical point of the network model then occurs for $t_{A}=t_{B}=\frac{1}{\sqrt{2}}$, and varying $\theta$ away from $\pi/4$ drives the system into either a spin insulator or a quantum spin Hall state, with a jump in the spin quantum Hall conductance at the transition.

The simplest way to define Green's functions is to use a first quantized formalism where the single particle wave function $\psi(r,t)$ follows a discrete time evolution $\psi(r,t+1) = U \psi(r,t)$ where the evolution operator $U$ is a sparse $2N \times 2N$ symplectic matrix for a network with $N$ links. $U$ simply describes the discrete evolution of the wave function with random $SU(2)$ matrices on each link and the correct scattering factor at each node. The retarded Green's function between two links $i$ and $j$ is then defined as
\begin{equation}
G(i,j)_{\sigma_i,\sigma_j } = \Braket{i,\sigma_i | (1-z U)^{-1} | j,\sigma_j},
\label{eqGreenSQHE}
\end{equation}
where $z=\mathrm{e}^{i \epsilon -\eta}$ with the energy $\epsilon=0$ at the transition and $\eta=0^+$ an infinitesimal level broadening. The advanced Green's function is defined by replacing $z$ by $z^{-1}$, but because the evolution operator is a symplectic matrix, it is possible to relate the advanced Green's functions in terms of the retarded one, so that all physical observables can be expressed in terms of the retarded Green's function~\eqref{eqGreenSQHE} only. At the transition~\cite{PhysRevLett.82.4524}, the Green's functions show a critical behavior, so that for example $\overline{{\rm tr} G(i,j) G(j,i)} \sim r^{-1/2}$. Note that the averaged local density of states is critical and scales as $\rho(\epsilon) \sim \epsilon^{1/7}$ at the transition~\cite{PhysRevLett.82.4524}, in sharp contrast with the IQHE case. 

\subsection{Second quantization and supersymmetry}

The SQHE network model can also be conveniently described within a second-quantized framework. One then thinks of the evolution operator as a product of transfer matrices, and introduces fermions $f^\dagger_{\sigma}$ to represent the retarded Green's function. Usually, one needs both retarded and advanced operators to describe the product of retarded and advanced observables, but as already described above, we will only need the retarded Green's function in the case of the SQHE effect. In order to average over disorder, we use either the replica or the supersymmetry trick. Within the replica trick, the Green's function can be represented as 
\begin{equation}
G(i,j)_{\sigma_i,\sigma_j }= \lim_{n \to 0} \langle f^a_{\sigma_i}(i) f_{\sigma_j}^{a \dagger}(j) \rangle,
\end{equation}
with, in the path integral formalism, 
\begin{equation}
\langle f^a_{\sigma_i}(i) f_{\sigma_j}^{a \dagger}(j) \rangle = \int {\cal D}[f,\overline{f}]  f^a_{\sigma_i}(i) \overline{f^{a}_{\sigma_j}}(j) {\rm e}^{z \sum_{b=1}^n \overline{f^{b}} U f^b},
\end{equation}
where $f^{a=1,\dots,n}={f^a_{\sigma_i}(i)}$ is a vector of size $2N$, and the path integral measure reads ${\cal D}[f,\overline{f}] = \prod_{a=1}^{n} \prod_{i,\sigma_i} d f_{\sigma_i}^a(i) d \overline{f_{\sigma_i}^{a}}(i) {\rm e}^{-  \overline{f_{\sigma_i}^{a}}(i)   f_{\sigma_i}^{a}(i) }$. Because of the replica limit $n \to 0$, the average over disorder can be realized explicitly and it is useful to think of the critical point of the disordered system (the SQHE plateau transition) as obtained from a limit $n \to 0$ of a family of CFTs with central charge $c(n) \to 0$ as $n \to 0$. 

\begin{figure*}
\centering
    \includegraphics[scale=0.45]
    {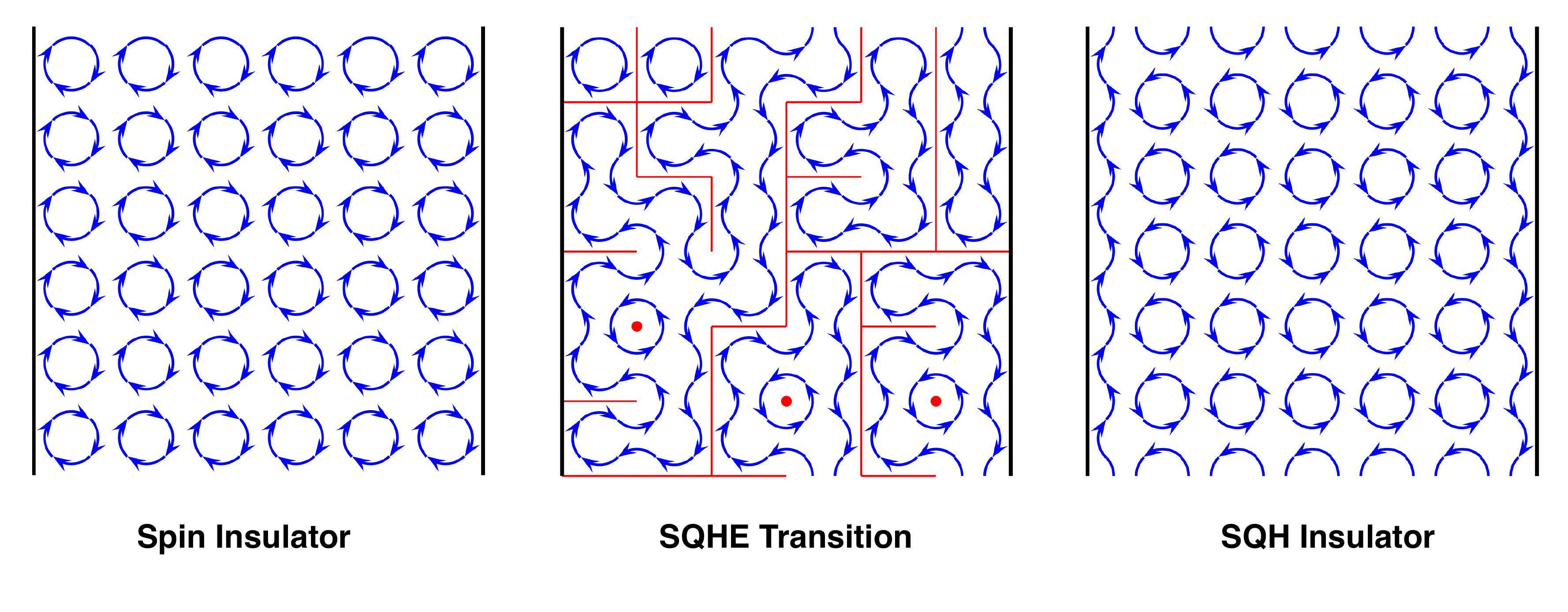}
    \caption{Percolation description of the SQHE transition~\cite{PhysRevLett.82.4524}, with periodic (resp. open) boundary conditions in the vertical (reps. horizontal) direction. The blue loops can be thought of as (semiclassical) paths contributing to spin transport, and their orientations correspond to alternating fundamental and dual representations of the superalgebra ${\rm sl}(2|1)$. Average over disorder configurations enforces that these loops be non-intersecting, so that they can be mapped onto hulls of classical percolation clusters (shown in red in the middle picture only). The SQHE transition then corresponds to the critical point of the percolation problem, separating a trivial spin insulator phase from a SQH insulator phase with a jump is the spin quantum Hall conductance. }
\label{FigSQH}
\end{figure*}

A more productive approach in the context of the SQHE is to use the so-called supersymmetry trick in order to average over disorder~\cite{PhysRevLett.82.4524} (see also Refs.~\onlinecite{PhysRevB.55.10593,Cardy:2005ab}). Instead of a family of fermions (or bosons) parametrized by $n \to 0$ as in the replica tick, one introduces two bosons $b_{\sigma}$ and two fermions $f_{\sigma}$ per link (one per spin), with canonical commutation relations on ``up links'' of the network, and modified anti-commutation relations for the fermions on down links (see Ref.~\onlinecite{PhysRevLett.82.4524} for details). The equal number of fermions and bosons ensures that closed loops in the graphical expansion of the partition function vanish, so that $Z_{\rm SUSY}=1$. In the path integral formalism~\cite{Cardy:2005ab}, this can be understood very easily as $Z_{\rm SUSY} =  \int {\cal D}[f,\overline{f}] {\cal D}[b,b^\dagger] {\rm e}^{z \overline{f} U f + z b^\dagger U b} \sim 1$, thanks to the Gaussian integral identities  $\int {\cal D}[f,\overline{f}] {\rm e}^{z \overline{f} U f }={\rm Det}(1-z U)=(\int {\cal D}[b,b^\dagger] {\rm e}^{z b^\dagger U b })^{-1}$. In the second quantized framework, one ends up with a transfer matrix that commutes with the superalgebra ${\rm osp}(2|2)$ (see {\it e.g.} Ref.~\onlinecite{1996hep.th7161F} for a review of Lie superalgebras) for any disorder realization, and the average over disorder then projects the Fock spaces on each up or down links onto $SU(2)$ singlets. The resulting Fock spaces can be shown to correspond to the fundamental representation of ${\rm sl}(2|1)\sim{\rm osp}(2|2)$ (for up links) and its dual (for down links), both with finite dimension 3 (super dimension $2|1$) with two bosonic states and one fermonic one. This drastic simplification -- the Fock spaces become finite-dimensional after averaging over disorder -- allows one to solve this supersymmetric model exactly by mapping it onto the classical percolation problem. We will not need the details of this mapping here and instead refer the interested reader to Ref.~\onlinecite{PhysRevLett.82.4524}, but we simply remark that using the isomorphism ${\rm osp}(2|2) \simeq {\rm sl}(2|1)$, the resulting loop models obtained by expanding graphically the partition function corresponds precisely to the supersymmetric ``$SU(2|1) \sim SU(n \to 1) \sim {\rm percolation}$'' model studied in Sec.~\ref{SecLoopModel} (see~\ref{secSUSYperco} in particular). This mapping is summarized in Fig.~\ref{FigSQH}. Note that the only SQHE observables that can be mapped onto classical percolation in this way can be described using two bosons $b_{\sigma}$ and two fermions $f_{\sigma}$ per link only. For instance, multi-fractal properties usually require more species of fermions and bosons, with supersymmetry ${\rm osp}(2m|2m)$ and $m > 1$, and this mapping onto percolation breaks down~\cite{0305-4470-36-12-323,PhysRevB.78.245105}.

\subsection{Logarithmic correlations from the replica trick}

\subsubsection{Observables acting on $N=1$ link}
\label{SecN1SQHE}

We now focus on the replica approach and start from fermions $f_\alpha=f_\sigma^a$, with $\alpha=(a,\sigma)$ and $a=1,\dots,n$, with $n$ the number of replicas. We work with fermionic operators because bosonic variables in Anderson transitions typically involve non-compact symmetries, leading to a much more complicated representation theory analysis. For symmetry class $C$, we expect the global symmetry of the network model to be $Sp(2n)$, the group of unitary matrices $U$ satisfying $U^T \sigma^y U=\sigma^y$. We start with observables of the form $f_\alpha(i) f^\dagger_\beta(i)$ where $i$ labels the links of the networks. The conjugate of the fundamental representation $[1]^\star$ of $Sp(2n)$ is pseudo-real $[1]^\star \simeq [1]$ since $U^* = \sigma^y U \sigma^y$. It is therefore convenient to introduce $c^a_\sigma = \sum_{\sigma^\prime} (i \sigma_y)_{\sigma \sigma^\prime} f^{a \dagger}_{\sigma^\prime}$ which transforms as $c \mapsto U c$ if $f \mapsto U f$ with $U \in Sp(2n)$. The combination $f^T \sigma^y c$ is then clearly invariant, and the $n(2n-1)-1$ antisymmetric operators
\begin{equation}
W_{\alpha \beta} = f_\alpha c_\beta - f_\beta c_\alpha  + \frac{\sigma^y_{\alpha \beta}}{n} f^T \sigma^y c,
\end{equation}
form an irreducible representation of $Sp(2n)$. The $n(2n+1)$ symmetric counterparts $V_{\alpha \beta} = f_\alpha c_\beta + f_\beta c_\alpha$ then correspond to the adjoint representation of $Sp(2n)$.
The average over disorder in the network model enforces the physical operators to be $SU(2)$ singlets. Among all these operators, we therefore restrict the physical observables to 
\begin{equation}
\phi_{ab}(i) = W_{a \uparrow b  \downarrow} = \sum_{\sigma} \left(f^a_\sigma f_\sigma^{b \dagger} - \frac{\delta_{ab}}{n} \sum_{c=1}^n f^c_\sigma f_\sigma^{c \dagger}  \right),
\label{eqPhiSQHE}
\end{equation}
which transform under the adjoint representation of the subgroup $U(n) \subset Sp(2n)$ (see Sec.~\ref{subsubsecReducedSym} below). 
We thus expect the two-point function of $\phi_{ab}$ for generic $n$ to take the form
\begin{equation}
\langle \phi_{aa}(r_i)\phi_{bb}(r_j) \rangle = A(n) \left(\delta_{ab} - \frac{1}{n }\right) r^{- 2\Delta_\phi(n)},
\label{eqPhiPhiSQHE}
\end{equation}
where the scaling dimension $\Delta_\phi(n)$ is obviously different from the one used in Sec.~\ref{SecLoopModel} for $SU(n)$ models, in particular, its general expression as a function of $n$ is not known. We also define the invariant operator $\Phi = \sum_\sigma \sum_{c=1}^n f^c_\sigma f_\sigma^{c \dagger}$, along with $\hat{\Phi} = \Phi - \langle \Phi \rangle$. The two-point function of $\hat{\Phi}$ is then given by $\langle \hat{\Phi}(r_i) \hat{\Phi}(r_j) \rangle = n B(n) r^{-2  \Delta_\Phi(n)}$ with $B(0) \neq 0$ and $\Delta_\Phi(n)$ is the scaling dimension of $\Phi$.

Disorder-averaged correlation functions in the network model are then obtained by considering the limit $n \to 0$, which is ill-defined in eq.~\eqref{eqPhiPhiSQHE}. The only way to fix this divergence is to require $A(0)=B(0)$ and $\Delta_\phi(0) = \Delta_\Phi(0)$, and introduce $\psi_a = \phi_{aa} +\frac{1}{n} \hat{\Phi}$. The $n \to 0$ limit of $\psi_a$ is then well-defined 
\begin{equation}
\langle \psi_{a}(r_i)\psi_{b}(r_j) \rangle = A(0)  r^{- 2 \Delta_\phi(0)} \left(\delta_{ab} + \kappa \log \frac{r}{a} \right),
\label{eqSQHEN1_Log}
\end{equation}
with $\kappa =2 \lim_{n \to 0} (\Delta_\phi -\Delta_\Phi)/n$. The same correlation function can also be expressed in terms of disorder-averaged Green's functions in the SQHE network problem using Wick's theorem
\begin{multline}
\langle \psi_{a}(r_i)\psi_{b}(r_j) \rangle = \overline{{\rm tr} G(i,i) {\rm tr} G(j,j)}  -\overline{{\rm tr} G(i,i)} \times \overline{ {\rm tr} G(j,j)}\\
- \delta_{ab} \overline{{\rm tr} G(i,j)G(j,i)},
\label{eqSQHEN1_Wick}
\end{multline}
where ${\rm tr}$ represents the trace over the spin index -- recall that the retarded Green's function $G$ is actually a $2 \times 2$ matrix. Comparing equations~\eqref{eqSQHEN1_Log} and~\eqref{eqSQHEN1_Wick}, we can then infer the scaling of various Green's functions as a function of the distance $r$.

As discussed above, using the supersymmetry mapping~\cite{PhysRevLett.82.4524}, one can then relate these Green's functions to percolation observables (see also Refs.~\onlinecite{PhysRevB.65.214301,Cardy:2005ab, 0305-4470-36-12-323, PhysRevB.78.245105}). For instance, it is  straightforward to show that $\overline{{\rm tr} G(i,i) {\rm tr} G(j,j)} = \sum_{\sigma_i, \sigma_j}\langle b_{\sigma_i}(i) b^\dagger_{\sigma_i}(i) f_{\sigma_j}(j) f^\dagger_{\sigma_j}(j) \rangle_{\rm SUSY}= \langle (1+2 B(i)) (1+2 Q_z(j)) \rangle_{\rm SUSY} =1$, where $B$ and $Q_z$ are two of the eight SUSY generators  of ${\rm sl}(2|1)$ in the fundamental representation (see {\it e.g.} Ref.~\onlinecite{1996hep.th7161F} or the appendix of Ref.~\onlinecite{PhysRevB.78.245105}) and we have used the fact that ${\rm str} B Q_z ={\rm str} B = {\rm str} Q_z = 0 $, with {\rm str} the supertrace in the fundamental representation. Therefore, $\overline{{\rm tr} G(i,i) {\rm tr} G(j,j)}  -\overline{{\rm tr} G(i,i)} \times \overline{ {\rm tr} G(j,j)}  =0$ so the correlation function that could show a logarithm at this order is trivially zero. Similarly, one finds that the non-logarithmic part $\overline{{\rm tr} G(i,j)G(j,i)}$ is given by $-2 {\mathbb P}^{(1)}_1(i,j)$ where ${\mathbb P}^{(1)}_1(i,j)$ is the probability -- introduced above in Sec.~\ref{SecLoopModel} -- that $i$ and $j$ belong to the same loop. This shows~\cite{PhysRevLett.82.4524} that $\overline{{\rm tr} G(i,j)G(j,i)}$ decays with the 2-leg (1-hull) exponent which allows us to identify $\Delta_\phi(n=0) = \frac{1}{4}$. These results imply that the parameter $\kappa$ in eq.~\eqref{eqSQHEN1_Log} is actually zero. Note that nothing in our symmetry analysis prevents this from happening. This could be just a coincidence, {\it i.e.} the derivative of the exponents $\Delta_{\phi}(n)$ and $\Delta_{\Phi}(n)$ with respect to $n$ could coincide at $n=0$, but this could also indicate the presence of a larger symmetry that enforces $\Delta_{\phi}(n)=\Delta_{\Phi}(n)$. We emphasize that the actual symmetry of the critical point of the SQHE is most likely must larger that $U(n)$ or $Sp(2n)$, so we always run into the risk of predicting logarithms that appear with amplitudes $\kappa=0$. This was also true for the loop models of Sec.~\ref{SecLoopModel} -- for which it is known that the actual symmetry of the critical point is much larger than $SU(n)$~\cite{Read2007263}, but there we had a way to compute $\kappa$ exactly, see eq.~\eqref{eqDelta}. Actually, had we considered the above $SU(n)$ model at a mean-field level (above the critical dimension), we would have also found $\kappa =0$ instead of~\eqref{eqDelta}. Our approach is nevertheless fully consistent -- the symmetry analysis correctly predicts that $\overline{{\rm tr} G(i,j)G(j,i)}$ should be a scaling operator -- and we deduce that observables acting on a single link do not have logarithmic correlations.

\subsubsection{$Sp(2n)$ vs $U(n)$ symmetry}
\label{subsubsecReducedSym}

In order to find logarithmic observables, we thus turn to operators acting on two nearest neighbor links $i$ and $i^\prime$ as in Fig.~\ref{Fig2pointObs}. Before doing so, we first remark that equation~\eqref{eqPhiSQHE} suggests restricting to $SU(2)$-invariants operators from the very beginning: $\sum_{\sigma} f^a_\sigma f_\sigma^{b \dagger}$, and analyzing the $U(n)$ symmetry of the remaining indices. Incidentally, the remaining symmetry $U(n)$ coincides with the denominator in the coset $Sp(2n)/U(n)$  of the sigma model describing the transition. This simplification is motivated by the fact that the genuine symmetry of the critical point is unknown  anyway, so that we choose to work with the subgroup $U(n)$ of $Sp(2n)$ in order to simplify dramatically the calculations.  As we will see in the following, this will turn out to be enough to identify observables that should have logarithmic correlation functions at the critical point. 

We remark that this is related to our choice of considering quadratic observables $f_\alpha(i) f^\dagger_\beta(i) $ from the beginning. It is of course quite natural starting from operators like $f_\alpha(i) f^\dagger_\beta(i) $ that the $SU(2)$ invariance has to be enforced within the replica approach {\sl before} taking the limit $n \to 0$ since the spin structure is completely lost in the replica limit $Sp(2n \to 0)$. A more correct approach would be to consider more complicated representations $V$ and $V^\star$ of $Sp(2n)$ acting on a single edge, that are compatible with the $SU(2)$ symmetry and that involve products of fermionic operators with different replica indices (see Ref.~\onlinecite{Affleck:1985aa} in the context of the IQHE). Other observables would then be constructed by taking tensor products of these representations. However, we will see in the following that some of the logarithmic structure of the SQHE can be understood by starting with the much simpler quadratic operators $\sum_{\sigma} f^a_\sigma f_\sigma^{b \dagger}$  satisfying the $SU(2)$ local (``gauge'') invariance,  classifying them using the remaining unitary symmetry of the replica indices and analyzing the limit $n \to 0$. We also note that in the SUSY language, this amounts to considering the much simpler subgroup $U(1|1)$ of $OSp(2|2)\sim U(2|1)$ whose indecomposable representations are suspected to play a crucial role in disordered systems~\cite{0305-4470-35-27-101}.

\subsubsection{Observables acting on $N=2$ links}

We now turn to observables acting on two nearest neighbor sites $i$ and $i^\prime$. Let us consider operators of the type $\sum_{ \sigma_i,\sigma^\prime_i} f^a_{\sigma_i}(i)f^{b \dagger}_{\sigma_i}(i)f^c_{\sigma_{i^\prime}}(i^\prime)f^{d\dagger}_{\sigma_{i^\prime}}(i^\prime)$, which are manifestly $SU(2)$ invariant. As in Section~\ref{SecLoopModel}, we expect observables symmetric under the exchange $a \leftrightarrow c$ or $b \leftrightarrow d$ to reproduce the physics of the operators acting on $N=1$ link, with no logarithm for $n=0$. We thus focus on the antisymmetric tensor $T_{abcd}(r_i)=-Q_{ab}(i)Q_{cd}(i^\prime)+Q_{ad}(i)Q_{cb}(i^\prime)-Q_{ab}(i^\prime)Q_{cd}(i)+Q_{ad}(i^\prime)Q_{cb}(i)$, with $Q_{ab}(i)= \sum_\sigma f_\sigma^a (i)f_\sigma^{b \dagger}(i)$. We then consider $\Phi^{(2)}=\sum_{ab} T_{aabb}$ which is invariant under $U(n)$, and introduce $\hat{\Phi}^{(2)}=\Phi^{(2)}-\langle \Phi^{(2)}\rangle$. The fields $\phi^{(2)}_{ab} = \sum_c T_{abcc}$ then transforms in the adjoint representation of $U(n)$. These two operators  $\hat{\Phi}^{(2)}$  and $\phi^{(2)}_{ab}$ have the same symmetry as $\Phi$ and $\phi_{ab}$ introduced in section~\ref{SecN1SQHE}, and the same mechanism leading to logarithms in the limit $n \to 0$ applies here as well. Because they have the same symmetry, the ill-defined limit $n\to0$ indicates logarithms in {\it subleading} contributions~\cite{Vasseur2014435}, so that we find, in the limit $n \to 0$ 
\begin{multline}
\langle \psi^{(2)}_{a}(r_i)\psi^{(2)}_{b}(r_j) \rangle = A \delta_{ab} r^{- 2 \Delta_\phi(0)} \\ +B r^{- 2 \Delta^{(2)}_\phi(0)} \left(\delta_{ab} + C \log \frac{r}{a} \right) + \dots ,
\label{eqSQHEN2_Log}
\end{multline}
where $\psi^{(2)}_a = \phi^{(2)}_{aa} +\frac{1}{n} \hat{\Phi}^{(2)}$. As we have seen above, $\Delta_\phi(0)=\frac{1}{4}$ corresponds  to the one-hull percolation exponent, while $ \Delta^{(2)}_\phi(0)$ is {\it a priori} unknown from this replica analysis, but we will shortly see that it is given by the two-hull percolation exponent,  $\Delta^{(2)}_\phi(0)=\frac{5}{4}$. 

To translate this into a concrete prediction for the network model, we evaluate the correlator $\langle \psi^{(2)}_{a}(r_i)\psi^{(2)}_{b}(r_j) \rangle$ in terms of disorder-averaged Green's functions. We find 16 different contributions which we compute using Wick's theorem, including for example
\begin{multline}
\lim_{n \to 0} \sum_{c,d=1}^n \sum_{\lbrace \sigma \rbrace } \langle f_{a,\sigma_1}(p_1) f^\dagger_{c,\sigma_1}(p_1) f_{c,\sigma_2}(p_2) f^\dagger_{a,\sigma_2}(p_2)  \\  \times f_{b,\sigma_3}(p_3) f^\dagger_{d,\sigma_3}(p_3) f_{d,\sigma_4}(p_4) f^\dagger_{b,\sigma_4}(p_4)\rangle \\
=\lim_{n \to 0} \sum_{c,d=1}^n \overline{\sum_{\lbrace \sigma \rbrace }  \underset{i,j=1,\dots,4}{\det} \left( G_{\sigma_i \sigma_j}(p_i, p_j) \delta_{\alpha_i \beta_j}\right)},
\end{multline}
with $\alpha \in \lbrace a,c,b,d \rbrace$, $\beta \in \lbrace c,a,d,b \rbrace$ and $p\in \lbrace i,i^\prime,j,j^\prime \rbrace$. Computing these 16 determinants explicitly and taking the limit $n \to 0$, we find an expression that can be recast as
\begin{multline}
\langle \psi_{a}(r_i)  \psi_{b}(r_j) \rangle = 4 \left(\Omega(r) - \Omega_\infty + \Xi(r) + \Theta(r) \right) \\
- 4 \delta_{ab} \left( 2 \times \Xi(r) +  \Theta(r) \right).
\label{eqSHQEWickEx}
\end{multline}
The most important piece of this correlator is the disconnected part
\begin{align}
\Omega(r)& = \overline{ {\rm tr} G(i,i) {\rm tr} G(i^\prime,i^\prime) {\rm tr} G(j,j){\rm tr} G(j^\prime,j^\prime)} \notag \\
&+\overline{ {\rm tr} \left[G(i,i^\prime) G(i^\prime,i)\right] {\rm tr} G(j,j){\rm tr} G(j^\prime,j^\prime)} \notag \\
&+\overline{{\rm tr} G(i,i) {\rm tr} G(i^\prime,i^\prime) {\rm tr} \left[G(j,j^\prime)G(j^\prime,j) \right]} \notag \\
&+\overline{ {\rm tr} \left[G(i,i^\prime) G(i^\prime,i)\right] {\rm tr} \left[G(j,j^\prime)G(j^\prime,j) \right]},
\label{defOmegaSQHE}
\end{align}
which has a finite limit as $r \to \infty$:
\begin{equation}
\Omega_\infty = \left( \overline{ {\rm tr} G(i,i) {\rm tr} G(i^\prime,i^\prime)}+  \overline{ {\rm tr} G(i,i^\prime)G(i^\prime,i)} \right)^2.
\end{equation}
The functions $\Theta(r)$ and $\Xi(r)$ can also be readily computed and contain 16 and 4 terms respectively, but their explicit expression will not be relevant to our purposes. Indeed, in analogy with the discussion in Sec.~\ref{SecLoopModel}, we expect the logarithmic correlations to arise from {\it disconnected} terms such as $\Omega(r)$, and not from connected observables like $\Theta(r)$ and $\Xi(r)$ -- at this point, this statement is a conjecture based on analogy but it will be verified below when we compare our results to the supersymmetric approach. Based on this and using eqs.~\eqref{eqSQHEN2_Log} and~\eqref{eqSHQEWickEx}, we thus obtain 
\begin{equation}
\Omega(r) -\Omega_\infty \sim \alpha r^{-1/2} + \beta r^{-5/2} \log \frac{r}{a} + \dots 
\label{eqLogSQHEOmega}
\end{equation}
with $a$ the lattice spacing (UV cutoff), where we have anticipated the fact that $\Delta^{(2)}_\phi(0)=\frac{5}{4}$ (see supersymmetric approach below). Even though~\eqref{eqLogSQHEOmega} is enough to show the existence of logarithmic correlations in the disordered SQHE network model (which would naively contradict scale invariance and therefore immediately implies the existence of logarithmic operators!), it may seem like a rather weak prediction since the logarithm appears only as a correction to a power-law function, which itself is a subdominant contribution. We will see in the following that the IQHE transition leads to a prediction that should be (much) more appropriate for numerical checks.

\subsection{Logarithmic correlations from the supersymmetry trick}

We now argue that the logarithmic correlation~\eqref{eqLogSQHEOmega} obtained from the replica trick can be recovered using an independent method based on the results of Sec.~\ref{SecLoopModel} and the SUSY description of some SQHE observables in terms of percolation probabilities~\cite{PhysRevLett.82.4524} (see also Refs.~\onlinecite{PhysRevB.65.214301,Cardy:2005ab, 0305-4470-36-12-323, PhysRevB.78.245105}). Recall that the SUSY description involves two bosons $b_{\sigma}$ and one fermion $f$ on each link, corresponding to the fundamental representation of the superalgebra sl($2|1$). In that language, eq.~\eqref{defOmegaSQHE} can be written as
\begin{equation}
\Omega(r) = \sum_{\lbrace \sigma \sigma^\prime \rbrace} \langle b_{\sigma_i}  b_{\sigma_i}(i)^\dagger b_{\sigma_{i^\prime}}  b_{\sigma_{i^\prime}}(i^\prime)^\dagger  b_{\sigma_{j}}  b_{\sigma_{j}}(j)^\dagger b_{\sigma_{j^\prime}} b_{\sigma_{j^\prime}}(j^\prime)^\dagger \rangle_{\rm d},
\label{eqSUSYOmega}
\end{equation}
where $r=|r_i-r_j|$, and the subscript $d$ in the correlator refers to the {\it disconnected} part (with a finite limit $r \to \infty$) involving Wick contractions between $i$ and $i^\prime$ or $j$ and $j^\prime$ only. This correlation function can easily be expressed in terms of the SUSY generator $B=\frac{1}{2}(b_\uparrow^\dagger b_\uparrow+b_\downarrow^\dagger b_\downarrow+1)$, so that it can be expanded onto various classical percolation probabilities, with amplitudes given by the supertrace of powers of $B$ in the fundamental representation (see {\it e.g.} Ref.~\onlinecite{PhysRevB.78.245105} for related calculations). Similarly, we find that $\Omega_\infty = \left( \overline{ {\rm tr} G(i,i) {\rm tr} G(i^\prime,i^\prime)}+  \overline{ {\rm tr} G(i,i^\prime)G(i^\prime,i)} \right)^2=(1+4 \langle B(i) B(i^\prime)\rangle_{\rm SUSY})^2= F_d^\infty $, where $ F_d^\infty= ({\mathbb P}^{(1)}_1(i,i^\prime)- {\mathbb P}^{(0)}_1(i,i^\prime))^2$ was introduced in Sec.~\ref{SecLoopModel}. Gathering these different pieces, we find the following expression in terms of percolation probabilities
\begin{equation}
\Omega(r) -\Omega_\infty \sim F_d(r)-  F_d^\infty,
\end{equation}
where $F_d(r)$ is precisely the linear combination of {\it disconnected} percolation probabilities introduced after eq.~\eqref{eqLogLoopLattice}. As we have argued in Sec.~\ref{subseclogPerco}, $F_d(r)$ has a leading power-law behavior associated with the 1-hull percolation exponent, with logarithmic corrections to the subleading term -- associated with the 2-hull or 4-leg percolation exponent -- that can be understood as the singular limit $n \to 1$ of $SU(n)$ loop models. Using these results, we thus conclude that eq.~\eqref{eqLogSQHEOmega} holds, in agreement with the replica analysis. Therefore, the logarithmic correlations in percolation described in Sec.~\ref{subseclogPerco} have the same physical origin as the logarithm in~\eqref{eqLogSQHEOmega}, and they correspond to the ``mixing'' of the energy and 2-hull (4-leg) operators in the 2D percolation problem. 

\section{(Tentative) Generalization to the IQHE transition}

\label{secIQHE}

We have shown above that both the replica trick and the supersymmetry trick -- combined with the analysis of $SU(n \to 1)$ loop models -- predict the existence of logarithmic correlations at the SQHE plateau transition, associated with concrete disorder-averaged observables (see eq.~\eqref{defOmegaSQHE}). Although the SUSY mapping onto percolation is obviously very specific to the SQHE transition, our analysis of the $n\to 0$ limit of the replica trick can be generalized to other symmetry classes. To illustrate this point, we conclude this paper by shortly describing the case of the Integer Quantum Hall Effect (IQHE), with symmetry class A. The network model describing the transition was introduced in the seminal work of Chalker and Coddington~\cite{0022-3719-21-14-008}, the only difference with the SQHE case being the random $U(1)$ phases -- instead of $SU(2)$ matrices -- on each link. The associated topological sigma model~\cite{Pruisken:1984aa} with bosonic fields reads
\begin{equation}
{\cal L} = \frac{\sigma_{xx}}{8} {\rm tr} (\partial_\mu Q)^2 - \frac{\sigma_{xy}}{8} \epsilon_{\mu \nu }{\rm tr}  Q \partial_\mu Q \partial_\nu Q, 
\end{equation}
where $Q$ lives on the coset $U(n,n)/U(n)\times U(n)$ (or $U(1,1|2)/U(1|1) \times U(1|1)$ in the SUSY formulation~\cite{,Weidenmuller:1987aa}), with $n$ the number of replicas, and $\sigma_{xy}$ plays the role of a topological angle associated with the non trivial second homotopy group $\pi_2 = {\mathbb Z}$ of the target space. At the plateau transition, the sigma model flows to strong coupling and we expect the quantum critical point to have a large symmetry, and we will assume that $U(n,n)$ is a subgroup of this symmetry. To avoid having to deal with non-compact symmetries, we restrict ourselves to the case of fermonic replicas, with symmetry group $U(2n)$. We thus write the retarded and advanced Green's functions as $G^\pm(i,j)= \lim_{n \to 0} \langle f^\pm_a(i) f_{a}^{\pm \dagger}(j) \rangle$, where $a=1,\dots,n$, with the $U(2n)$ symmetry acting on the indices $\alpha=(a,\pm)$.

\subsection{Observables acting on $N=1$ link}

Using the symmetry analysis of  Sec.~\ref{SecLoopModel}, we introduce the operators $\phi_{\alpha \beta} = f_\alpha(i) f^\dagger_{\beta}(i) - \frac{\delta_{\alpha \beta}}{n} \sum_\gamma f_\gamma(i) f^\dagger_{\gamma}(i)$, which could potentially lead to logarithms in the limit $n \to 0$ because of the second term. However, the average over disorder in the Chalker-Coddington model leads to observables with the same number of retarded and advanced particles on a given link -- this can be understood in a graphical expansion as retarded paths come with a random phase ${\rm e}^{i \gamma}$ while retarded paths are weighted with ${\rm e}^{-i \gamma}$, integrating over $\gamma$ then restricts the number of paths of each types to be the same. We therefore consider the observables $\phi_{ab}(i)=f^\pm_a(i) f_{b}^{\mp \dagger}(i)$ which transform irreducibly under the reduced symmetry $U(n)\times U(n)$. Note that this is similar to the reasoning that led us to consider a reduced $U(n)$ symmetry for the SQHE transition in order to ensure that all observables were $SU(2)$ singlets, a condition enforced by the average over disorder. Once again we emphasize that this is related to our choice to start with simple quadratic operators: the vertex model corresponding to the $U(2n)/U(n)\times U(n)$ sigma model is in fact built out of more complicated representations $V$ and $V^\star$ of $U(2n)$ (involving products of up to $n$ fermion operators) compatible with the local $U(1)$ invariance~\cite{Affleck:1985aa}. As in the SQHE case, we choose to focus instead on the much simpler operators $\phi_{ab}(i)=f^\pm_a(i) f_{b}^{\mp \dagger}(i)$ and to restrict our analysis to the subgroup $U(n)\times U(n)$ of $U(2n)$. (From the SUSY point of view, this means that we are considering only some part of the indecomposability of the supergroup $U(1,1|2)$ corresponding to the much simpler subgroup $U(1|1) \times U(1|1)$.) We expect this analysis to be enough to uncover potentially logarithmic observables as $n \to 0$.

There is no $1/n$ pole in correlation functions at this order (this differs from the SQHE case) and the only non-trivial correlator is $\langle \phi_{aa}(i) \phi_{bb}^\dagger(j)\rangle$
\begin{equation}
\lim_{n \to 0} \langle f^\pm_a(i) f_{a}^{\mp \dagger}(i) f^\mp_b(j) f_{b}^{\pm \dagger}(j) \rangle = -\delta_{ab} \overline{ G^\pm(i,j) G^\mp(j,i)},
\end{equation}
consistent with the fact that $\overline{ G^\pm(i,j) G^\mp(j,i)}$ should be a scaling operator.  

\subsection{Observables acting on $N=2$ links}

As for the SQHE, one needs to consider coarse grained observables acting on more than one link to generate logarithmic correlations. We thus consider products of operators acting on a single link of the form $ f^\pm_a(i) f_{b}^{\mp \dagger}(i) f^\mp_c(i^\prime) f_{d}^{\pm \dagger}(i^\prime) $, which can be thought of as being obtained from `fusing' $N=1$ operators -- we recall that $i$ and $i^\prime$ are close neighbors. These $n^4$ operators form a reducible representation of $U(n)\times U(n)$. There is a unique invariant $T^{(0,0)} = \sum_{ab} f^+_a(i) f_{b}^{- \dagger}(i) f^-_b(i^\prime) f_{a}^{+ \dagger}(i^\prime) = {\rm Tr} Q^{+ -}(i) Q^{- +}(i^\prime)$, where the trace symbol corresponds to a sum over the indices and $Q_{ab}^{+-}=f^+_a(i) f_{b}^{- \dagger}(i)$. There are also two representations corresponding to the products trivial $\times$ adjoint or  adjoint $\times$ trivial in retarded/advanced spaces, given by $T_{ab}^{(0,1)} = \sum_{c} f^+_c(i) f_{a}^{- \dagger}(i) f^-_b(i^\prime) f_{c}^{+ \dagger}(i^\prime) - \frac{\delta_{ab}}{n} T^{(0,0)}$ and $T^{(1,0)} = \sum_{c} f^+_a(i) f_{c}^{- \dagger}(i) f^-_c(i^\prime) f_{b}^{+ \dagger}(i^\prime) - \frac{\delta_{ab}}{n} T^{(0,0)}$. Because of the $1/n$ terms in these operators, these operators could potentially have an ill-defined $n \to 0$ limit that would eventually lead to logarithmic correlations. However, it is not hard to see using Wick's theorem that because of the sum in the first terms, the  $n \to 0$ limit of, say, the correlator $\langle T_{aa}^{(1,0)}(i,i^\prime) T_{bb}^{(1,0)\dagger}(j,j^\prime)\rangle$ is well defined. In other words, the correlation function $\langle T_{aa}^{(1,0)}(i,i^\prime) T_{bb}^{(1,0)\dagger}(j,j^\prime)\rangle$ will involve a term $\delta_{ab}-\frac{1}{n}$ dictated by representation theory that, similarly to other examples we have studied in this paper, could lead to logarithms in the limit $n \to 0$; but it appears with an overall ${\cal O}(n)$ pre-factor that cancels the $1/n$ pole. After subtracting the invariant and these two representations, we are left with $n^4-1-2\times(n^2-1)=(n^2-1)^2$ operators that transform under the product of adjoint representations under $U(n)\times U(n)$. These operators are given by
\begin{multline}
\psi_{abcd} = f^+_a(i) f_{b}^{- \dagger}(i) f^-_c(i^\prime) f_{d}^{+ \dagger}(i^\prime) \\- \frac{\delta_{bc}}{n}  \sum_{k}f^+_a(i) f_{k}^{- \dagger}(i) f^-_k(i^\prime) f_{d}^{+ \dagger}(i^\prime)\\ - \frac{\delta_{ad}}{n}  \sum_{k}f^+_k(i) f_{b}^{- \dagger}(i) f^-_c(i^\prime) f_{k}^{+ \dagger}(i^\prime) + \frac{\delta_{bc}\delta_{ad}}{n^2} T^0.
\end{multline}
We thus expect the two-point function $\langle \psi_{abba}  \psi^\dagger_{cddc} \rangle$ for generic $n\neq 0$ to scale algebraically as $r^{-2 \Delta_{\psi}(n)}$ with an amplitude proportional to $(\delta_{ad}-\frac{1}{n})(\delta_{bc}-\frac{1}{n})$. Similarly to the previous examples encountered in this paper, we introduce a new operator $\hat{\psi}_{abba}=\tilde{\psi}_{abba}- \langle \tilde{\psi} \rangle$ and $\tilde{\psi}_{abba}= f^+_a(i) f_{b}^{- \dagger}(i) f^-_b(i^\prime) f_{a}^{+ \dagger}(i^\prime) =\psi_{abba}+\frac{1}{n}(T^{(1,0)}_{aa}+T^{(0,1)}_{bb})+\frac{1}{n^2} T^0$ in order to solve the ill-defined limit $n \to 0$. Because there are three different operators $\psi_{abba}$, $T^{1,0} \sim T^{0,1}$ and $T^{0}$ involved, the resolution of the `$n \to 0$ catastrophe' is more intricate than the other examples encountered in this paper\footnote{In particular, there is the possibility of a mixing of these three operators into a rank-3 Jordan cell for the scale transformation generator, leading to $(\log r)^2$ terms in correlation functions. }. We leave the detailed understanding of this $n \to 0$ limit for future work, but simply notice that regardless of these details, the $n \to 0$ limit will yield $\langle \hat{\psi}_{abba}  \hat{\psi}^\dagger_{cddc} \rangle \sim r^{-2 \Delta_\psi(0)}  \log r $ if $a \neq d$ and $b \neq c$, with the possibility of having $\log^2 r$ terms as well. The important point is that in analogy with the other examples described in this paper, we expect to have logarithmic corrections in the {\it disconnected} part of the correlation function  $\langle \hat{\psi}_{abba}  \hat{\psi}^\dagger_{cddc} \rangle$, obtained by enforcing all the replica indices $a \neq d$ and $b \neq c$ to be different. Using Wick's theorem, we find
\begin{equation}
\langle \hat{\psi}_{abba}  \hat{\psi}^\dagger_{cddc} \rangle = \Omega(r=|r_i-r_j|)-\Omega_\infty, 
\end{equation}
for $a \neq d$ and $b \neq c$ with
\begin{equation}
\Omega(r)= \overline{ G^+(i,i^\prime) G^-(i^\prime,i) G^+(j,j^\prime) G^-(j^\prime,j)}.
\label{defOmegaIQHE}
\end{equation}
The function $\Omega(r)$ has a (non-universal) finite limit as $r \to \infty$, given by $\Omega_\infty = \overline{ G^+(i,i^\prime) G^-(i^\prime,i) }^2$ which is independent of $r_i$ -- recall that $i$ and $i^\prime$ are chosen to be in the infinitesimal neighborhood of $r_i$. The replica approach thus predicts
\begin{equation}
\Omega(r) - \Omega_\infty \sim r^{-2 \Delta_\psi(0)}  \log^\alpha r,  
\label{eqLogIQHEOmega}
\end{equation}
with $\alpha=1$ or $\alpha=2$. Contrary to the SQHE case~\eqref{eqLogSQHEOmega}, logarithms appear in the leading power-law contribution, and there is no need to subtract a very complicated combination of Green's functions to isolate the logarithmic term. The downside is that the limit $n\to 0$ is less controlled in the IQHE case as it involves three operators, leading to the indetermination of the exponent $\alpha=1$ or $\alpha=2$ in~\eqref{eqLogIQHEOmega}. Note also that the term coming with amplitude $\delta_{ad} \delta_{bc}$ in the correlation function $\langle \hat{\psi}_{abba}  \hat{\psi}^\dagger_{cddc} \rangle$ should scale purely algebraically, so that 
\begin{equation}
\Gamma(r) = \overline{ G^+(j,i) G^+(i^\prime,j^\prime) G^-(i,j) G^-(j^\prime,i^\prime)}   \sim r^{-2 \Delta_\psi(0)}.
\end{equation}
Gathering these different pieces, we conclude that
\begin{equation}
\frac{\Omega(r) - \Omega_\infty}{\Gamma(r)} = \frac{\overline{ G^+(i,i^\prime) G^-(i^\prime,i) G^+(j,j^\prime) G^-(j^\prime,j)}- \Omega_\infty}{\overline{ G^+(j,i) G^+(i^\prime,j^\prime) G^-(i,j) G^-(j^\prime,i^\prime)}} ,
\label{eqLogOnly}
\end{equation}
should scale purely logarithmically at the plateau transition, either as $\log r$ or as $\log^2 r$. It would be really interesting to compute this quantity numerically in the Chalker-Coddington model and to try to fit it with $\alpha \log^2 r + \beta \log r + \gamma$ to check this prediction.

Expanding Green's functions graphically in terms of retarded and advanced paths~\cite{PhysRevB.86.165324,PhysRevB.84.144201}, these logarithmic correlations appear from the mixing of the operator creating two advanced paths and two retarded paths (which is essentially the analog of a watermelon operator, with dimension $X_{2,2}$ in the notations of Ref.~\onlinecite{PhysRevB.84.144201}), and the operator $T^{(0,0)}(r_i)$ which is just counting the number of retarded and advanced paths around the neighborhood of $r_i$, corresponding to the ``thermal''-like perturbation driving the system out of criticality by taking $z \neq 1$. 

\subsection{Remarks on open vs closed quantum networks and conductance correlations}

We emphasize that our field theory predictions apply to {\it closed} quantum networks, for which no lead is connected to the system. In particular, quantities like $\overline{G^+(i,j)G^-(j,i)}$ (diffusion propagator) in terms of the Green's functions~\eqref{eqGreenSQHE} require an infrared regulator in order to be finite (an infinitesimal level broadening making $\left| z\right|$ slightly less than 1). From a numerical point of view, it is more convenient and efficient to work with {\it open} quantum networks that are also more natural in the context of transport. For example, to define the point contact conductance $g(i,j)$ , one simply cuts in half two links $i$ and $j$ (the contacts) with each cut leading to one in-going link (source of current) and one out-going link (drain). This amounts to attaching two leads to the system. The conductance can then be computed within the Landauer--B\"uttiker formalism in terms of the transmission matrix between the leads (see Ref.~\onlinecite{PhysRevB.59.15836} for details). Unfortunately, opening links in the network correspond to the insertion of additional operators that make our analysis more complicated~\cite{PhysRevLett.112.186803}. Moreover, the point-contact conductance $g(i,j)$ is in general completely unrelated to $G^+(i,j)G^-(j,i)$ (except in the SQHE case where their average are actually identical up to a constant) -- see in particular the discussion in Ref.~\onlinecite{0305-4470-36-12-323}. It is nevertheless tempting to conjecture that the symmetry considerations of this paper extend to $g(i,j)$ as well, even if in general it scales in a different way as $G^+(i,j)G^-(j,i)$.  This conjecture is motivated by the naive guess that the open or closed nature of the network model should not affect our symmetry-based analysis: it is then natural to expect logarithms as well in observables involving $g(i,j)$ instead of  $G^+(i,j)G^-(j,i)$. The analog of eq.~\eqref{defOmegaIQHE} in terms of conductances would give an observable characterizing the correlations between the local transport properties (conductances between $i$ and $i^\prime$, and between $j$ and $j^\prime$) of two remote regions $r_i$ and $r_j$. It would be very interesting to investigate numerically (or theoretically using a different approach) whether such physically appealing observables show logarithmic correlations at criticality.

\section{Discussion}

\label{secConclusion}

In this paper, we have generalized the ideas of Cardy~\cite{1999cond.mat.11024C} to analyze the physical origin of logarithmic correlations in Quantum Hall plateau transitions using the replica trick. In the case of the Spin Quantum Hall transition, we found a combination of Green's functions that scales logarithmically on average at the critical point (see eqs.~\eqref{defOmegaSQHE} and~\eqref{eqLogSQHEOmega}). Alternatively, we recovered this result independently using the percolation description of the SQHE transition~\cite{PhysRevLett.82.4524} and a replica-like analysis of the $n \to 1$ limit of $SU(n)$ dense loop models. Using a similar argument for the Integer Quantum Hall transition, we uncovered a relatively simple observable that should have logarithmic correlations at the plateau transition (see eqs.~\eqref{defOmegaIQHE} and~\eqref{eqLogIQHEOmega}).

We emphasize that these predictions based on the replica limit of a global symmetry group at the critical point should be taken with care: as we have already mentioned repeatedly throughout this manuscript, considering a symmetry group smaller than the actual symmetry of the critical point may in principle completely spoil our predictions: different representations for the smaller group that are mixed and lead to logarithms in the replica limit $n \to 0$ could in fact be part of the same larger irreducible representation for the actual symmetry group of the system. In other words, the amplitudes in front of the logarithms within our approach could in principle be zero. However, we recall that the very same approach lead to results in very good agreement with numerical simulations~\cite{1742-5468-2012-07-L07001} and with more rigorous algebraic approaches~\cite{2014arXiv1409.0167G} in simpler critical points including the two dimensional percolation problem studied in Sec.~\ref{SecLoopModel}, even though the actual symmetry there is also much larger than $SU(n)$ as well~\cite{Read2007263}. Moreover, the replica approach seems to lead to very natural predictions for the scaling operators of the critical theory, with results that are consistent using two different approaches (replica trick {\it vs} supersymmetry) in the SQHE case. 

Our predictions should therefore be taken as reasonable conjectures as to where to look for logarithmic correlations in quantum Hall plateau transitions. It would obviously be very important to verify the results of this replica analysis numerically, and we expect in particular that the prediction~\eqref{eqLogOnly} should be very useful in that respect, since it should scale purely logarithmically at the critical point, with no power-law dominant contribution. Our field theory predictions apply naturally to closed quantum networks, and it would be crucial to see to what extent they generalize to open networks that are more natural from the point of view of transport, and that seem more practical numerically. We note that we considered simplified observables for the SQHE and the IQHE transitions (see~\ref{subsubsecReducedSym} above): we leave the detailed analysis of the $Sp(2n)$ or $U(2n)$  symmetry of these theories for future work. It would also be very interesting to investigate to what extent the replica and the supersymmetry approaches agree for the SQHE transition: for example, the replica analysis provides a natural expression for the ``watermelon'' operators scaling with the $k$-hull percolation exponents. Whether these expressions agree with the mapping onto percolation {\it via} the supersymmetry trick remains unknown. We expect that pushing further this symmetry-based replica analysis of Quantum Hall transitions should be very helpful to make progress towards a deeper understanding of the CFTs describing these critical points.

 {\noindent\bf Acknowledgements.} I wish to thank J. Cardy for useful discussions that motivated this work and for helpful comments on the manuscript. I also thank J.L. Jacobsen and H. Saleur for collaborations on related matters and J.E. Moore, A. Nahum, A.C. Potter and especially R. Bondesan for insightful discussions and comments on the manuscript. I am supported through the Quantum Materials program of LBNL.

\bibliographystyle{mybibstyle}

\bibliography{BibQuantumHall}
    
    \end{document}